# Theoretical Aspects of Immunity

Michael W. Deem and Pooya Hejazi
Rice University
6500 Main Street – MS 142
Houston, TX 77030
mwdeem@rice.edu

## Abstract

The immune system recognizes a myriad of invading pathogens and their toxic products. It does so with a finite repertoire of antibodies and T cell receptors. We here describe theories that quantify the immune system dynamics. We describe how the immune system recognizes antigens by searching the large space of receptor molecules. We consider in some detail the theories that quantify the immune response to influenza and dengue fever. We review theoretical descriptions of the complementary evolution of pathogens that occurs in response to immune system pressure. Methods including bioinformatics, molecular simulation, random energy models, and quantum field theory contribute to a theoretical understanding of aspects of immunity.

## Table of Contents

## 1. Introduction

Just as in statistical mechanics there is a search through configuration space to find regions of favorable free energy, so too in evolution there is a search through sequence space to find proteins, pathways, and organisms of high fitness. The immune system is a real-time example of evolution, as our body creates and modifies molecules of the immune system in response to invading pathogens. Many pathogens respond to this immune pressure by evolving themselves. The present article is a brief appraisal of the literature in theoretical aspects of the adaptive, vertebrate immune system.

The immune system exhibits many of the mathematical features that make biology interesting (1). Finite timescales in biology bring out the importance of dynamics, and finite population sizes bring out the importance of randomness, correlations, and fluctuations. Biological information and structure is organized hierarchically, and modularity is a key principle that we will use in describing the immune system. At a deeper level, the emergence of the adaptive immune system in jawed vertebrates several hundred million years ago is an example of spontaneous emergence of modularity, a symmetry breaking phase transition by which biology itself emerged from organic chemistry several billion years ago (2, 3). Finally, the response of the immune system induces a complementary evolutionary dynamics in pathogens.

In this chapter, we discuss some of the theoretical aspects of immunity. In section 2, we provide an overview of the adaptive immune system, including a discussion of the generalized NK theory that has been developed to describe immune system dynamics. In section 3, we cover traditional ODE-type models of the immune system. In section 4, we describe some of the bioinformatic approaches to vaccine design. We continue the discussion of vaccine design with a consideration of theories of the collective, population-level effects in section 5. We elaborate with an explicit consideration of the dynamics of virus spread in section 6. We explain the statistical mechanics of the immune response to influenza in section 7. We present the statistical mechanics of the immune response to dengue fever in section 8. The pathogen reacts to the immune response, and we describe general theories of pathogen evolution, some derived from quantum field theory, in section 9. We discuss the theories used to understand the evolution of influenza virus in section 10. We describe theories for other diseases in section 11. We conclude with a very brief sample of theories to understand the immune response to cancer in section 12.

## 2. Brief summary of the immune system

Our immune system protects us against death by infection. There are two major components of the adaptive vertebrate immune system—the antibody-mediated humoral response and the T cell mediated cellular response. In this article, we discuss quantitative models of the immune system response to often highly variable pathogens. The variation exists either because the disease mutates at a non-negligible rate or because the disease exists in multiple subtype forms.

These models can help deduce design rules for vaccines. The finite size of the immune system repertoires and the existence of immune system memory make the immune system response to variable viral diseases nontrivial. A number of reviews have covered some of the interesting aspects and challenges the immune system poses (4-7).

### 2.1 The humoral, B cell immune response

Virus, bacteria, and other pathogens are recognized and repelled from infecting our cells by antibodies, proteins produced by our B cells. Antibodies recognize patches, called antigens, on the surface of pathogenic proteins. See Figure 1. To recognize the many different pathogens that can attack us, we have a repertoire of different B cells. Each different B cell produces a distinct antibody. The human immune system has on the order of $10^8$ different B cells, and the probability that a given antigen will be recognized by a given antibody is about 1 in $10^5$.

The immune system uses a hierarchical strategy to search through the space of possible antibodies to find those which recognize antigens. There are about 100 amino acids in the variable region of an antibody chain, and so there are about $20^{100}$ or $10^{130}$ possible such structures. The immune system searches this large space with the sparse set of $10^8$ structures available via a two step process. The first step is the initial creation of B cells that produce antibodies by joining together three different antibody fragments. There are libraries of these fragments available in the genome of the individual, and this diversity can create on the order of $10^{11}$ different structures, although only $10^8$ are present at one point in time. This assembly process is called VDJ recombination. When an individual is challenged by a pathogen, the B cells that produce the 1 in $10^5$ subset of these initial antibodies that recognize the antigen are further optimized by a local search process. Essentially, the DNA that encodes these antibodies is randomly mutated, and the cells that make the antibodies that better recognize the antigen are expanded in concentration and number. This mutation and selection process is repeated for a number of rounds and leads to modified antibodies that strongly recognize the antigens of the pathogen. This second, optimization process is called somatic hypermutation.

**Figure 1:** Influenza virus hemagglutinin (Aichi 1968) trimer complex with a neutralizing antibody dimmer (PDB accession number 1E08).

While the immune system does not evolve the single, optimal antibody to recognize an antigen, it does evolve a subset of antibodies that recognize the antigen. The number of distinct antibodies produced in high concentrations is on the order of a half-dozen or so. Indeed, the antibody response can typically be deconvoluted to responses from a few specific antibodies (8). These antibodies typically bind to non-overlapping neutralizing epitopes on the pathogen. This set of epitopes on the surface of pathogenic proteins is termed the "protective immunome."

The portion of the antibody that binds the antigen, which is also the region that undergoes somatic hypermutation, is called the hypervariable region. The antibody protein structure in the hypervariable region is constrained to be loop-like, largely independent of sequence. Fundamental computer simulations have verified this aspect of antibody structure (9). Interestingly, simple molecular dynamics was insufficient to equilibrate these structures, and hybrid Monte Carlo with multi-dimensional parallel tempering (10) was required. Parallel tempering simulations show that the hypervariable region, while constrained to be loop-like, is sufficiently flexible to bind a wide range of antigens (9, 11).

### 2.1.1 The generalized NK (GNK) model

The dynamics of affinity maturation can be described theoretically as a search through sequence space for antibodies with increased binding constants to antigen. The immune system response to an antigen generates a high concentration of only a few memory B cells in response to a pathogen. This focusing of the immune diversity results in a strong subsequent response to identical antigens, but can reduce the effectiveness of a response to different but related antigens. This competition between memory sequences and newly formed VDJ recombinations determines the effectiveness of a vaccine against a virus.

The Deem group has developed a random energy theory that describes the dynamics of the immune system and the interaction between influenza antigens and antibodies (12). This theory takes into account both the ruggedness of the interaction energy landscape upon which the antibodies evolve and the correlations in this landscape that allow VDJ recombination to produce a viable naive repertoire. The model is population based, considering a population of B cells in one person, a population of viruses, and a population of individual people. The random energy model permits study of the immune response at the level of individual antibodies and antigens. Absent this theory, study of the molecular interactions in detail would be computationally prohibitive, since there are many influenza strains, $6.7 \times 10^9$ people, $10^8$ antibodies per individual , and $10^4$ atoms per antibody. Use of random energy theory to treat correlations in complicated physical systems has a long pedigree in statistical physics.

The generalized NK (GNK) model describes the interaction between antibodies and antigens. Each antibody is identified by its amino acid sequence, and each antigen is represented by the coupling parameters in the theory. A successful immune response to an antigen corresponds in the theory to finding a set of amino acid sequences with low energy for the coupling parameters that represent the antigen. The spin-glass form of the GNK model makes the energy landscape rugged. The energy landscape has a structure, however, which is induced by the local antibody structure ($U_{sd}$), the secondary antibody structure ($U_{sd\text{-}sd}$), and the interaction with the influenza proteins ($U_c$) (2).

$$U = \sum_{i=1}^{M} U_{\alpha i}^{sd} + \sum_{i>j=1}^{M} U_{ij}^{sd-sd} + \sum_{i=1}^{P} U_{i}^{c}$$

An important feature of the theory is that the rugged landscape upon which the B cells that code for antibodies evolve is not completely random, but rather has structure. This structure is what allows productive VDJ recombinants to be formed. We were motivated to use a spin-glass, or random energy model, for the fitness landscape because we knew that evolution of proteins by point mutation alone is a slow process and hence the landscape must not be smooth. Moreover, we realized that the random energy model must be correlated, because proteins have secondary structures at an intermediate length scale between the amino acid and the domain length scales. We used this theory to develop new ‘subdomain swapping’ protocols for protein evolution (13, 14).

The approach is innovative because the GNK model can predict immunodominance and cross reactivity. The GNK model considers the full diversity of the B or T cell response, and this is why it can predict phenomena such as immunodominance, cross reactivity, and original antigenic sin (see Sections 2.1.3, 7.1, 7.3, and 8.1). Traditional ODE models can be fit to, but cannot predict, immunodominance, cross reactivity, and vaccine efficacy data. Traditional ODE models consider the immune dynamics within a small number of classes of immune cells, e.g. naive, activated, and memory, but do not consider the full diversity of $10^8$ B or T cells that exist within the human immune system. The GNK theory considers the full $10^8$ diversity and for this reason is able to predict the immune dynamics and vaccine efficacy that result from skewing of and interactions among this diversity.

### 2.1.2 Scaling theory of the B cell immune response

A physically-motivated, scaling theory of the immune system gives us some insight into the principles of immune recognition (15). Larger animals have more mass to protect, and so they have a larger immune system. In fact, both the diversity and the total number of B cells is greater in larger animals. Scaling theory suggests how these quantities depend on the mass, M, of an animal.

The scaling theory starts with the observation that the lifespan, $T_0$, of an animal scales with body mass as $T_0 \sim M^{1/4}$ (15). The immune system must be sufficiently protective that the animal survives pathogen attack for this period of time, on average. The mass of the animal is composed of $\sim M^{3/4}$ service volumes, the volume of tissue supplied and drained by one capillary. Each service volume is of volume $\sim M^{1/4}$, and it is independently monitored by the immune system. The capillaries that supply blood flow to these service volumes are of universal size and function in all animals, and so the amount of blood and antigens that enter the service unit per unit of time is independent of M. If one assumes that the mobility of the B cells is also independent of M, then the time required for one B cell that is specific for one antigen to search each unit scales as the volume, $\sim M^{1/4}$. To keep the time of detection constant with animal size, there must be $\sim M^{1/4}$ copies of each B cells in each of the units. Thus, the total number of copies of each B cell should scale as $\sim M^{1/4} \times M^{3/4} \sim M$.

The second observation of scaling theory is that the total number of infections of an individual is proportional to the lifetime metabolic intake of the animal (15). The total metabolic rate of an animal scales as $\sim M^{3/4}$ according to Kleiber's law. The lifetime scales as $\sim M^{1/4}$, and so the lifetime metabolic intake $\sim M$. Thus, the number of infections against which the immune system must protect is c M, where c is some constant. The probability that an immune system with N different B cells will fail to recognize a specific pathogen is $\varepsilon = (1 - V_0/V)^N \sim \exp(-N V_0/V)$ where $V_0/V$ is the fraction of antigens recognized by a specific antibody. The probability that all lifetime infections are successfully repelled is $(1 - \varepsilon)^{cM} \sim \exp(-\varepsilon c M)$. This probability should be near unity, and so $\varepsilon c M \ll 1$. Using the expression for $\varepsilon$, we find $N \gg (V/V_0) \ln(c M)$.

This prediction that the number of distinct B cells scales as $\sim \ln(c M)$, with $\sim M$ copies of each B cell, is consistent with known diversities in humans and mice (15). For example, humans have roughly $10^8$ different B cells, with $10^5$ copies of each. Mice, which weigh about 20g, roughly 3400x smaller in mass than humans, have about $10^7$ different B cells with 10 copies of each. At small enough masses, this scaling relation must break down. Using the numbers for human and mice, we find $c = 0.12$ / g. Thus, the scaling relation surely breaks down for $c M < 1$, or $M < 8$ g. Recent estimates of the immune repertoire size in zebrafish suggest a B cell diversity of 5000 – 6000 (16). Zebrafish, however, with an average mass of 0.3 g, would not seem to be within the scaling regime.

### 2.1.3 Autoimmune disease and the functional role of glassy dynamics in the immune system

The VDJ and somatic hypermutation hierarchy has evolved to allow the immune system to rapidly recognize a wide variety of potential antigens. One might ask, however, if even more efficient evolutionary dynamics might be possible in the immune system, and if so why such dynamics has not evolved into wide usage. In fact, if the immune system continued with recombination during the somatic hypermutation process, this would be more efficient dynamics (17, 18). In other words, antibodies that more strongly recognize antigens could be found with this process. It turns out, however, that these antibodies would bind not only their intended target, but also other targets. See Figure 2. This off-target activity would lead to autoimmune disease. It would appear, therefore, that the slow, glassy dynamics of evolution in the immune system serves a functional role to inhibit autoimmune disease.

This calculation also provides us with an estimate of how many antibodies might be needed to cover antigen space. As will be discussed in more detail below, there is a natural measure of antigenic distance in sequence space, which we term $p_{epitope}$. The immune system primed by exposure to an antigen appears able to recognize other, related antigens out to a distance of $p_{epitope} = 0.19$ (19) to 0.45 (20). Taking the larger number, we find that such an immune response recognizes about

$$N(p) = 19^{i} \times 20!/[i! \times (20-i)!] \cong 10^{-9}$$

of all possible antigens, where $i = 20 \times p_{epitope}$ (17). Thus, an immune repertoire with the diversity to allow the generation of $10^9$ memory antibodies, not necessarily all at the same time, would be able to recognize most antigens. This upper limit agrees well with the $10^8$ naïve diversity of the human antibody immune system.

**Figure 2:** Cross reactivity of primary immune response memory antibody sequences for the two strategies against altered antigens (point mutation only PM, and gene segment swapping in addition to point mutation PM + GSS), where $K$ is the binding constant and $p$ the antigenic distance of the new altered antigen from the original. The cross reactivity ceases at a higher value ($p > 0.472$) for PM + GSS in comparison with PM alone ($p > 0.368$). From (17).

## 2.2 The cellular, T cell immune response

**Figure 3:** 3-D representation of TCR-MHC-peptide complex (PDB accession number 2CKB). A) Mouse TCR bound to the class I MHC H-2Kb molecule and peptide – backbone tube diagram of ternary complex. B) Mouse TCR α and β chains binding the MHC-peptide complex – above view of CDR regions. C) Above view of surface region of the class I MHC H-2Kb molecule and peptide. After (21).

The cellular immune system performs a stochastic search of T cell receptors (TCRs) to recognize antigenic peptide ligands that are presented by the MHC complex of individual cells (20). See Figure 3. Multiple identical TCRs on the cell membrane binding to the ligand activates the T cells, which like antibodies are constructed from modular elements, with each individual human having a diversity of approximately $2 \times 10^7$ different receptors (22).

T cells originate in bone marrow and mature in the thymus. They acquire their diversity through the stochastic process of VDJ recombination. During the development of TCRs, they undergo rounds of selection for increased avidity. Unlike antibodies, they do not undergo somatic hypermutation, presumably because further evolution may yield TCRs with unnecessarily high affinity and cross-reactivity against other short peptides present in the body and cause autoimmune disease (17, 18). Some mature T cells proliferate and produce effector T cells, whereas others become memory cells.

The important components of the T cell response are the peptide-MHCI complex (pMHC) and the T cell receptor (TCR). Typically, for $CD8^+$ T cells, the peptide is on the order of 9 amino acids long, and the T cell-mediated response consists of cycles of concentration expansion and selection for favorable binding constants. The replication rate of T cells in the immune system is a function of the fitness given by the generalized NK model. The binding energy quantifies activation of the T cells, and specific lysis quantifies the rate of killing infected cells. The replication rate defines replication, and the binding energy quantifies activation. Some T cells are activated more than others. T cell activation depends on their stimulation by antigen, as well as other stimulatory factors within the immune system. See Figure 4. The difference in replication rate between different T cells, as it depends on their stimulation by antigen, is of interest in the theory of immunity, and this difference depends on the amino acid identity of the TCR. Moreover, the replication rate changes when the peptide/MHCI complex to which the T cell binds changes. There are interactions within a subdomain of the TCR ($U_{sd}$), interactions between subdomains of the TCR ($U_{sd-sd}$), interactions between the TCR and the peptide ($U_{pep-sd}$), and direct binding interaction between the TCR and peptide ($U_c$), as in Section 2.1.1 (23).

We use the GNK model to describe the interaction between the TCR and the pMHC, where the binding constant, K, is given by (23):

$$K = e^{a - bU}$$

The ability of one T cell to recognize an infected cell, z, is obtained by:

$$z = \frac{1}{N_{size}} \sum_{i=1}^{N_{size}} \min\left(1, K_i / 10^6\right)$$

The specific lysis, L, or the probability that the infected cell is lysed by the T cell, can be calculated using:

$$L = \frac{zE/T}{1 + zE/T}$$

**Figure 4:** The GNK model of the TCR selection dynamics where antigen recognition expands T cell concentration according to their level of stimulation by the antigen. Over 10 cell divisions the concentration can expand by three orders of magnitude, reducing the diversity to 0.5% in the primary response. When a previously infected individual with memory cells to epitope A is subsequently exposed to a new epitope B, the immune response shall be a combination of the memory response from the prior exposure to A and a naïve response to B. Our theory measures the contributions of both components. From (23).

Another general theory of the T cell response is kinetic proofreading, which seems to occur (24). The basic idea is that a series of binding interactions leading to activation can give a non-linear amplification of the signal. Denoting the complex between the T cell and the pMHC as $C_0$, a series of intermediate activated complexes is denoted by $C_i$, and the final active form is denoted $C_N$. Each activated complex satisfies

$$d\, C_i / d\, t = k\, C_{i-1} - k\, C_i - k_{-1}\, C_i, \qquad C_{N+1} = 0$$

Thus, one finds that the ratio of final activated complex to total complex is $C_N / \sum_i C_i = [k/(k+k_{-1})]^N$. Thus, the amount of activated complex depends in a highly sensitive and non-linear way on the equilibrium constant for T cell and peptide/MHC binding. In this way, ligands of different affinity may elicit qualitatively different signals.

A final theory of the T cell response is serial triggering (25). The mechanism of T cell activation seems to require multiple interactions with the TCR, not merely a single, very high affinity interaction. Indeed, if the affinity is too great, for example due to a low-off rate, then the TCR is not able to achieve a sufficient number of interactions. In this way, the T cell has evolved into a sensitive detector of small numbers of antigenic determinants present on infected cells.

### 2.2.1 Self-immunity

T cells created in the bone marrow pass through the thymus before proceeding to the lymph nodes and blood (26). Those T cells that bind no antigen in the thymus are eliminated. Those T cells that bind antigen weakly are kept. Those T cells that bind antigens from ones own proteins strongly are eliminated, because they would cause cross-reactivity and autoimmune disorders. Thus, T cell receptors are enhanced in weak-binding amino acids. Crystal structures support this conclusion that T cell recognition is via multiple, modest interactions. We can understand this suppression of strongly binding amino acids with the following physical argument (26). The interaction between the T cell receptor and the jth of the M peptides in the thymus is denoted by $-E(l,j)$, and the cutoff to avoid negative selection is denoted by $-E_N$. Thus, whether the T cell will survive negative selection, i.e. not bind too strongly to any of the peptides, is given by $p(l) = \prod_{j=1}^{M} \theta[(E_N - E(l,j)]$. The average survival probability is $<p(l)> \approx < \theta(E_N - E(l,j)>^M$, where one has assumed that the peptide compositions are independent. Making the further assumption, which we know from the GNK model is not true but which perhaps does not greatly detract from the present argument, that the interaction energy between the T cell receptor and the peptide is a linear sum of terms due to each amino acid on the peptide, $E(l,j) = \sum_{i=1}^{N} J(l_i,j)$, we reach the conclusion that $\sum_{i=1}^{N} J(l_i,j) < E_N$ for each of the j peptides. In other words, the distribution of amino acids in the T cell receptor must be such that $<J(l_i)> < E_N/N$. Thus, since the peptides in the thymus are typical self peptides, and the T cell receptor must bind them only weekly, the amino acids in the T cell receptor should be the weakly binding amino acids.

### 2.2.2 C-SMAC

The T cell receptor is bound to the surface of T cells. In some cases, the receptors aggregate during a recognition event, localizing the receptor, antigen, and kinases into a small area (27). This aggregation is now thought to regulate the degradation of the T cell receptors. A maximum of T cell signaling seems to occur for intermediate TCR-pMHC half lives. Too strong a binding does not allow serial triggering to cause activation and too weak a binding does not allow enough receptor binding to cause activation.

Advances in video microscopy revealed interactions between the T cell and antigen in lymph nodes (28). Three phases of T cell motion were found. The initial activation phase is termed phase I. Quantitative analysis showed that the duration of phase I, which can be minutes to hours, depends on the ratio of TCR-pMHC half life to the time of T cell – DC encounter (29). This ratio is a "consolidated measure of antigenic quality and type" (29).

## 3. Traditional ODE models of immune response and viral growth

Traditionally, theoretical modeling of the immune system has been based upon systems of ordinary differential equations (ODEs) that describe the time-dependent concentration profiles of pathogen, antigen, and B and T cells. A well studied case is that of HIV dynamics by the Los Alamos group (30). There are three phases of the disease: initial growth of the virus due to exposure and then suppression of the virus by the immune system, a long period of low virus concentration in the body, and then finally growth of the virus again with a transition to AIDS. Mathematical modeling along with experiment revealed rather rapid virus and immune cell turnover during the second phase, which had been previously thought to be a quiescent phase.

A subset of virus progeny are defective interfering (DI) particles. These particles may simply be failed attempts at viral replication. These DI particles may also serve a functional role to inhibit the immune response against the viable virus particles. A difference equation approach was used to evaluate the effect of DI particles on vesicular stomatitis virus (VSV) growth (31). At low DI particle concentrations, both virus and DI particles were predicted to expand in number. At intermediate concentrations, both were inhibited. Finally, at high DI particle concentrations, DI particle growth was inhibited, but virus was propagated.

A natural extension of ODE models is into the spatial dimension. Reaction diffusion models have been developed, for example, to examine the spatio-temporal patterns of viral infection (32). The model described VSV growth and spread in cells, using parameters extracted from imaging studies.

## 4. Bioinformatics of vaccine design

### 4.1 Modification of virus growth rates

The explosion of bioinformatics data, coupled with a mechanistic understanding of viral growth dynamics and immune recognition, allows for a novel approach to vaccine design. For example, a model-based approach has been applied to the design of growth-attenuated viruses (33). These are viruses to be used as a vaccine that generate an immune response but which grow poorly or not at all in humans and do not cause infection. The traditional experimental approach for creating such vaccines has been to evolve them in foreign hosts, until they replicate poorly and are no longer infectious. Using molecular biology and bioinformatics, and predicting how genetic changes correlate with virus growth, it is possible to design a growth-attenuated virus strain for use as a vaccine. A model predicted how a virus-infected cell will produce viral progeny. In particular, the model captures the details of virus entry, transcription of viral mRNA, translation of viral protein, replication of the viral genome, assembly of intermediates, and production and release of viral progeny (33). The predictions of this model were then used to suggest genetic modifications that would lead to a growth-attenuated strain.

In another study, the effect of gene order on virus growth was predicted and used to design a vaccine strain (34). Permutations of gene order for the vesicular stomatitis virus were examined for their effect on growth. Reordering of the genes that affect levels of gene transcripts by 6000× was found. These modifications of growth rates were largely due to imbalances in gene expression levels needed for optimal growth.

### 4.2 Epitope recognition

Bioinformatics methods have also been used to identify and modulate immune recognition of viruses. The Los Alamos group has extensively examined the humoral immune response to HIV. In one study, the B cell response to a DNA vaccine for HIV was examined (35). Nucleotide motifs that were common in HIV but under-represented in highly expressed human genes were identified. It was postulated that these motifs were leading to the poor expression of the HIV proteins gag, pol, and env from the vaccine. In addition, an optimized consensus codon sequence was chosen for the gag gene. These bioinformatics-based optimizations gave a 5× increased antibody response to the gag protein from the vaccine.

## 5. Collective effects of vaccine design in a population

Viruses exist not just in an isolated individual, but also in a population of individuals. Mathematical models can predict the sometimes non-intuitive effects of vaccine use in a population (36). A sufficiently powerful model would aid the effort to design a vaccine that is most effective in a population. One of the basic concepts in vaccine design is that of “herd immunity.” In the simplest model, viruses infect people and propagate from infected people to other susceptible individuals. There is a percolation transition as a function of the transmission probability of the virus. That is, no epidemic occurs below a critical transmission probability, whereas an epidemic that infects a significant fraction of the population occurs above the critical transmission probability. Under such a scenario the dynamical mean-field reaction rate equation is (37):

$$\partial_t \rho_k(t) = -\rho_k(t) + \lambda k[1 - \rho_k(t)]\Theta(\lambda)$$

where $\rho_k$ is the probability of a vertex with degree $k$ being infected, and $\Theta(\lambda)$ is the probability that any given link points to the infected node. At steady state, one obtains:

$$\rho_k = \frac{k\lambda\Theta(\lambda)}{1 + k\lambda\Theta(\lambda)}$$

Furthermore, we can obtain the epidemic threshold and the fraction of infective vertices, assuming a distribution that has finite degree fluctuations (37):

$$\lambda_c = \frac{< k >}{< k^2 >}$$

When $\lambda$ falls below $\lambda_c$, there is no epidemic. Vaccines affect the transmission probability, $\lambda$, because vaccinated individuals are much less likely to be infected by the virus. Thus, even people who are not vaccinated can benefit from the vaccine: if enough of the population is vaccinated and the vaccine is efficacious enough that $\lambda < \lambda_c$, they are completely protected. Decreasing the transmission rate below the critical value by means of vaccination so that no epidemic occurs is the essence of what is meant by "herd immunity." Recent exact mathematical results provide sharper bounds on the applicability of this mean field analysis to infinite populations (119).

A number of non-trivial issues arise in vaccine design against viruses with natural diversity (36). For example, different host species may be infected by different versions or "serotypes" or a virus. How should a vaccine composition be optimized if only some of the species will be vaccinated? The bacterial disease *H. influenza* (Hib) contains a single serotype whereas there are numerous serotypes of pneumococcal bacteria. Thus, the pneumococcal vaccine must be updated regularly, where as the Hib need not be. A related question for pathogens with multiple serotypes is which serotypes to include in the vaccine. Mathematical models provide insight into this question. For example, it might be advisable to include only the most pathogenic subtypes in the vaccine and leave out the avirulent serotypes. This approach, similar to the sparing of pesticides in a limited fraction of cultivated land, may allow the avirulent serotype to outcompete and suppress the more troublesome, virulent serotypes.

## 6. Epidemiology

Viruses exist in a population of individuals and survive by propagation between infected individuals. We can calculate an upper limit on the viral spreading rate (38). Let us assume that a single virus particle in a single cell reproduces in one day to make 100 viable progeny. The volume of a cell is $10^{-15}$ m$^3$. If each of the progeny infects one other cell, the volume of infected cells is $100^n \times 10^{-15}$ m$^3$ at the end of n days. This expansion process would be able to infect enough cells to populate the entire top km of the surface of the earth, a volume of $5.1 \times 10^{17}$ $m^3$, in 17 days.

We know that typical viruses do not overtake the globe quite this rapidly (38). The reason is that the population infected by the virus is heterogeneous. Living matter is not well mixed. Microbes, fish, plants, birds, and mammals are distributed heterogeneously on the earth, and virus progeny do not have unfettered access to all other susceptible cells on the earth. Virus transmission is highly restricted by the local contact processes between infected and susceptible individuals. In addition, viruses are restricted by their biology to the hosts they can infect. That is, a given virus usually can only infect a subset of cells in one, or at most a few, species.

Epidemiology is the study of virus spread in a population, and modern models take into account the heterogeneity of the transmission process and the selectivity of the virus. Mean field analysis of these models leads to classical results, but it fails for spatially localized contact processes. Approaches such as moment closure methods have been developed to account for correlations in the dynamics (39). High dimensional, detailed dynamics of an epidemic have been projected onto lower dimensions by projection operator methods (40). Some generic implications of spatial heterogeneity have been derived. For example, spatial heterogeneity of a host population tends to lead to greater disease persistence (41). Important order parameters to characterize viral spreading include how well the immune system recognizes related strains of a virus, transmissibility of the virus, and population size (42). Three fixed points for viral dynamics have been described: global extinction, stable single-strain persistence, and multiple-strain persistence with strain diversity. The transmission rate of the virus and strength of cross immunity between strains determine the boundary between single and multiple strain persistence. Among the order parameters, population size and cross immunity most dramatically affect the dynamics. It was noted that “The most crucial feature of multi-strain pathogeneiss is that infection by one strain induces partial immunity to future infections by other strains” (42). We will return to this point in Section 7.1.

Influenza provides an example of how the host affects viral transmission and strain structure. There is a worldwide effort to sequence human influenza, so a great deal is known about the distribution of the virus and its various strains. Human influenza A viruses tend to travel around the world with the international air travel of human hosts (43). Frequent travelers accelerate the spread of epidemics (44), although seasonal epidemics are large enough and infect enough people that even the average travelers are enough to distribute the virus globally.

Influenza predominates in the winter months of temperate climates. There are a number of suggested mechanisms to explain this observation (45). It may be that increased crowding that occurs with indoors activities, including school terms, during the winter months leads to greater transmission. The immune system may also be fundamentally weaker in cold weather. For example, the antibody response seems to depend on melatonin and be sensitive to vitamin D, both of which are lower during winter. Epidemics are most synchronized, i.e. more mean-field, in the more populous states of the US and during the most severe disease season, as one would expect from simple fluctuation versus number considerations. While children are still thought to be the dominant spreaders of disease at local levels of communities, schools, and households, the long-distance spread of the virus is better correlated with patterns of adult workflow traffic. Interestingly, epidemics in the tropics are correlated with rainy seasons. It is known, as well, that the virus is more stable at lower temperatures.

## 7. The immune response to influenza

Influenza is a viral disease that infects 5 – 15% of the population and causes a worldwide mortality of 250,000 – 500,000 annually. Influenza virus is categorized as either A or B. Influenza A is further subtyped by the hemagglutinin and neuraminidase serotypes, HxNy. The typical course of an influenza infection in an individual lasts 7 days (46). The first 1 – 2 days after exposure, the individual suffers no symptoms, although the virus is rapidly expanding in number in the upper respiratory tract. During these days, the individual is shedding virus and is infectious, however. Day 2 is the day of maximum virus shedding, which lasts roughly until day 7. The degree of infectiousness is roughly proportional to viral shedding. While fever is often considered a classic symptom of the flu, only 40% of A/H3N2, 36% of A/H1N1, and 7.5% of B cases cause fever. The H3N2 serotype is the most virulent and expands to higher viral concentrations, or titer, in individuals. H3N2 also typically causes more severe illness and higher mortality and a greater number of hospitalization cases than does H1N1. Children shed influenza virus earlier and longer than do adults.

There are multiple strains of each virus serotype. See Figure 5. The traditional method for quantifying difference between strains of influenza virus is known as the hemagglutination inhibition assay, which quantifies the ability of an antibody to competitively inhibit the binding of the hemagglutin protein of the virus to red blood cells. Antibodies raised against one strain of the virus typically inhibit other strains to lesser degree, in a way that depends on the difference between the strains. The lesser the inhibition, the greater the distance between the strains, and this assay forms a basis to quantify antigenic distance (47). These assays are typically performed in ferrets, however, and the correlation of antigenic distances derived from this animal model with vaccine efficacy in humans is imperfect.

The annual influenza epidemics exhibit a strong seasonality in temperate regions (48). It appears that epidemics are often associated with rainy seasons in tropical regions. Indeed, there is a correlation with vapor pressure, influenza transmission, and virus survival. It has also been observed that transmission decreases with vapor pressure in a linear way, and that virus survival increases with low levels of vapor pressure. Thus, it has been suggested that humidification of air in homes, schools, and work could reduce virus transmission substantially. From Section 5, we know that epidemics are percolation phase transitions, and so reduction of transmission probability below the critical value would extinguish an epidemic.

From sequence data, it is now thought that global migration of influenza drives epidemics (49). That is, the virus does not persist locally, flaring up during the winter. Rather, the virus seems to arrive from elsewhere. It is now thought that Asia, sometimes China and sometimes other places in Asia, is the source for the annual epidemic strains (50). For example, it appears that Northern hemisphere strains are not descendant of South American or African strains, but rather from Asian strains. This continual source of strains induces a one-dimensional character to the phylogenetic tree of influenza strains. Modeling efforts have suggested that once an influenza epidemic reaches $10^4$ – $10^5$ cases, travel restrictions are unable to contain further spread (51).

**Figure 5:** Phylogenetic tree of HA1 nucleotide sequences. From (50).

### 7.1 H3N2 influenza virus vaccine efficacy

There is an annual vaccine for influenza. There are three components to the vaccine, currently specific strains of A/H1N1, A/H3N2, and influenza B. In 2009/2010 there will also be a supplemental novel A/H1N1(2009) strain.

The immune response to an influenza vaccine and then subsequent response to a virus that may be a bit different from the vaccine has been the subject of intense theoretical efforts (19, 52, 53). The immune system recognizes mainly the hemagglutinin protein on the surface of the influenza virus particle. On this protein there are five main regions, called epitopes, that antiobodies recognize. It has been postulated that a suitable order parameter to characterize antigenic distance for influenza is the fraction of amino acids that differ in the dominant epitope between two strains (19, 53). This order parameter is called:

$$p_{\text{epitope}} = \frac{\text{(Number of amino acids that differ in the dominant epitope)}}{\text{(Number of amino acids in the dominant epitope)}}$$

This order parameter correlates to a greater degree with vaccine efficacy in humans than do even the ferret animal model studies (19, 52). This order parameter is a useful new tool for influenza vaccine design. See Figure 6.

**Figure 6:** Vaccine efficacy for influenza is well represented as a function of $p_{\text{epitope}}$. Also shown is a linear least square fit of the epidemiological data based on the $p_{\text{epitope}}$ parameter. Original antigenic sin, or negative efficacy, has been observed 26% of the time. From (19).

### 7.2 H1N1 influenza virus vaccine efficacy

The other major serotype of influenza that has afflicted humans over the past few decades, and the cause of some concern during the 2009/2010 season, is H1N1. As with H3N2, it is desired to have a predictive correlate of vaccine efficacy for H1N1. To develop the $p_{\text{epitope}}$ method, first the epitopes of H1N1 needed to be identified (54). Roughly 1/3 of the positions were known from antibody mapping experiments.

Remaining positions were identified by structural mapping from H3N2 and identification of sites under immune selection pressure. With these epitopes in hand, $p_{epitope}$ for H1N1 was developed (20). This order parameter correlated well with vaccine efficacy in humans, just as it did for H3N2.

An interesting distinction between H3N2 and H1N1 is that the immune response to H1N1 appears to be stronger (20). This, for example, explains why the H1N1 vaccines are typically more efficacious and why they extend protection to larger values of $p_{epitope}$ than do H3N2 vaccines. It also appears that due to this greater immune pressure, H1N1 viruses evolve to a greater extent when they are dominant than do H3N2 viruses.

### 7.3 Sequence space localization in the immune response to influenza

These measures of antigenic distance are based upon a statistical physics description of the immune response to vaccine and subsequent exposure to virus (2). This theory is termed the generalized NK (GNK) model. The GNK model is a spin glass, random energy theory of the landscape upon which the immune system dynamics evolves. This theoretical description was the first to predict both a region of positive protection from the vaccine and a region of increased susceptibility due to a vaccine. Essentially, there is an intermediate antigenic distance for which the immune memory generated by the vaccine causes the immune response to be trapped in a local region of sequence space, and this suboptimal response is worse than if the vaccine had not been given. For vaccine design, a positive efficacy is desired. The GNK theory allows one to quantify the antigenic distance between the vaccine and mutant viral strains and to predict the expected vaccine efficacy as a function of this distance.

## 8. The immune response to dengue fever

Originally a tropical and subtropical disease, Dengue virus (DENV) is now being tracked and investigated in the United States as well as the rest of the world (55, 56). The infected cases seem to grow with the range of the spread of DENV causing approximately 50 to 100 million cases on annual basis, with a conservative estimate of 25,000 mortalities each year (57). Despite being immune to the same strain after surviving a bout of the disease, individuals that are reinfected with a different serotype are at an elevated risk of suffering the dengue hemorrhagic fever (DHF) as a symptom. DHF is characterized by plasma loss due to increased vascular permeability and is the cuase for essentially all the mortalities stemming from DENV (58, 59). In order to avoid this “original antigenic sin” a DENV vaccine must be able to protect against all four serotypes simultaneously (60). To date, however, no four-component vaccine has been developed to combat all of the four serotypes (61). Development is impeded by the need to overcome, or at least minimize, immunodominance and bypass the potentially adverse effects caused by heterologous immunity.

There are four serotypes of dengue virus (62). The serotypes differ by up to 30% in amino acid composition. The ten DENV protein products, ordered by greatest to least variation, are NS2A, C, NS1, NS2B, E, NS4A, NS5, M, NS3, and NS4B. See Figure 7.

These four serotypes frequently co-circulate in the same locality. As with influenza, it appears that Asia provides the source population for dengue, and the greatest diversity of dengue is found in that region. In other words, there is frequent dengue virus migration, with relatively little in situ evolution (63). While the sequence data for dengue is not nearly as extensive as that for influenza, it does appear that there is significant fluctuation in genetic diversity of the virus, with rapid creation and extinction of many dengue clades.

## 8.1 Immunodominance in the immune response to dengue fever

Since DENV is comprised of four related serotypes, an ideal vaccine would provide the basis for a simultaneous and balanced attack against all four viral variants. Discovery of an effective vaccine against all four dengue viruses (DENVs), however, has been hampered by skewing of immune responses to only one or two serotypes. A fundamental problem with DENV is that immunity after infection by one serotype of DENV protects modestly or even negatively against reinfection by other serotypes (56, 57, 64). Presence of original antigenic sin requires having a vaccine for DENV that can induce protective immunity against all four serotypes simultaneously (56, 57, 65). To date, not only has no such effective vaccine has been developed, but also it is believed that the T cell immunological response to each serotype is to a limited number of epitopes that are similar but distinct in various serotypes (66, 67). Differences in the epitope sequences, in addition to simultaneous exposure to all four DENV serotypes can make the quality of the immune response to some of the serotypes poor (60, 65). Immunodominance is a severe immunological problem that DENV poses.

**Figure 7:** DENV genome expressed as proteins and the levels of antibody and T cell immune response to each (blank meaning no response, + indicating mild response, and +++ indicating the strongest response). From (57).

Clinical trials for a four-component DENV vaccine (57, 60) show an immunodominance effect, in which immune response is not strong against all serotypes. Why immunodominance occurs so dramatically in DENV has been the subject of much debate. Studies suggest that $CD8^+$ T cell receptors (TCRs) which recognize dominant epitopes inhibit expansion of TCRs to other epitopes, due to reduction of viral load, apoptosis, homeostasis, and resource competition (66, 68, 69). However, the complexity of the interactions between the immune system and DENV had made a quantitative understanding of immunodominance challenging. Moreover, no four-component vaccine to date has been able to overcome this immunodominance.

## 8.2 Multi-site vaccination to alleviate imunodominance in dengue fever

Based on our studies (23), we have suggested a novel multi-site vaccination that may ameliorate immunodominance for the DENV vaccine. See Figure 8. By multi-site vaccination we mean vaccinating in different locations in a manner that each component of the vaccine would drain to a physiologically distinct lymph node, e.g., as used to be done with the rabies vaccine in the abdomen and buttocks. See Figure 9. Hence there would be independent evolutions of immune response in different individual lymph nodes of various vaccine components until complete “mixing,” after which amplified T cells spread more evenly through the lymphatic system and a fraction are randomly chosen to be further expanded. Such independent selection evolution allows for a successful multi-site vaccination by generating a predisposition in amplified T cells prior to “mixing.” Since this predisposition can be preselected by vaccine components, it can in turn be used towards sculpting the immune response to ameliorate immunodominance. Administration of two different vaccines at physically separated sites rather than same site has been shown in cancer vaccines to reduce immunodominance (70-72). The model further suggests that vaccinating with a poorly recognized serotype first, followed by multi-site vaccination, is an effective strategy to sculpt an increased number of TCRs recognizing the subdominant serotypes to mitigate immunodominance. Finally, the model predicts that judicious choice of the subdominant epitopes may further ameliorate immunodominance.

**Figure 8:** Specific lysis ratios from the least to most dominant epitope of the four dengue viruses under various situations. Data from (23).

**Figure 9:** Representation of values for parameter mixing day to draining lymph nodes at different distances from the heart. From (21).

Three experimental studies in animals have now confirmed our hypothesis that multi-site vaccination may alleviate immunodominance (73-75). These studies investigated the diversity of a CD8 T cell response to a mixture of HIV epitopes. In (75), mice were immunized with a mixture of AL11 and KV9 $D_b$-restricted HIV epitopes. Injection to the same site resulted in a specific response to the KV9 epitope. Anatomic separation between injection sites resulted in a response against both epitopes. In (74), whether a broad CD8 T cell response recognizing multiple HIV-1 types (termed “clades”) could be induced by a multi-component vaccine was assessed in mice. Single-clade A, B, and C vaccines generated limited cross-clade reactivity. Combining the three clades into one vaccine resulted in a reduced breadth of response due to immunodominance. Administering individual clade-specific vaccines simultaneously into anatomically distinct sites on the body alleviated immunodominance and increased the number of epitopes recognized by the T cell response. In (73), a four-component dengue vaccine was examined in monkeys. A broader immune response was generated with multi-site vaccination than with single-site vaccination. A broader immune response was also generated when subdominant serotypes were administered in different physiological locations to the dominant serotypes. These studies provide confirmation of our theory.

## 9. General considerations of pathogen evolution

Emergence of drug resistant strains of pathogens exposes people to virulent infection and remains a significant problem (1). For example, A significant fraction of patients in hospitals are infected with common bacteria such as *Staphylococcus aureus*, *Enterococci*, and *Pseudomonas aeruginosa* resistant to antibiotics. Many species of viruses evolve in response to immune or vaccine pressure, and influenza has become resistant to several common antivirals.

### 9.1 The quasispecies theory of virus evolution

In the late 60s and early 70s, Crow, Kimura, and Eigen developed the quasispecies theory to describe viral evolution. These physical theories of evolution consider the replication and mutation of a population of viruses. The key results that emerged were that a population of viruses clusters around a defined genotype, which need not be the genotype with the most rapid replication rate, and that there is a critical mutation rate, beyond which the virus species no longer can exist. The disappearance of the virus for mutation rates greater than the critical value is called the error catastrophe.

There is an analogy between this evolutionary dynamics and thermodynamics: Individual viral fitness is like energy, mutation rate is like temperature, and fitness of the viral population is like free energy. Quantum field theoretical treatments were able to make this analogy precise (76). For example, if the viral replication rate, $f(\xi)$, depends on distance from a "wild-type" sequence, $\xi$, the population of viruses in the Crow-Kimura model optimizes the expression

$$f_m = \max_{\xi} \{f(\xi) - \mu + \mu[1 - \xi^2]^{1/2} \}$$

where $\mu$ is the viral mutation rate and $f_m$ is the mean viral fitness in the population. The error catastrophe is a true phase transition, with an order that depends on the form of the fitness function and the alphabet size of the virus (77). These methods provide the exact phase diagram and population fitness values of quasispecies theory, derivable from this equation.

Extension of quasispecies theory to multiple fitness peaks was possible by this quantum field theory (78). This extension allows one to consider, for example, one region of sequence space with high viral fitness and a distinct region of immune or drug suppression. This refined model allows one to consider evolution of drug resistance or immune escape of the virus.

The quantum field theory also allowed the incorporation of horizontal gene transfer (79) and recombination (80) to quasispecies theory. Horizontal gene transfer and recombination are essential for large-scale evolution, and an accurate description of viral evolutionary dynamics must include these processes. It was found that these processes affect the phase diagram and population fitness of the virus. Perhaps most significantly, however, these processes affect the ability of a finite population of viruses to evolve.

The quasispecies theory in the literature suggests using mutagens as antivirals. If the mutation rate of the virus could be driven beyond the critical value, then the virus would be eradicated. Some current antivirals may indeed operate by this mechanism. It has been shown in the case of vesicular stomatitis virus that the virus can evolve to decrease its natural mutation rate, thus counteracting this treatment strategy (81). This response to lethal mutagenesis is a novel viral resistance mechanism. Modulation of the mutation rate in response to a change in environmental pressure is well supported by generalized NK studies of evolution (13).

### 9.2 Co-evolution theory

The combination of the immune system and the virus population is a co-evolving system. In the viral dynamics, there is the error catastrophe phase transition as a function of viral mutation rate. In addition, however, there is an adaptation catastrophe for virus mutation rates too low to escape immune attack (82). Thus, viruses can only exist within a window of mutation rates. It might be expected that the immune system would try to make this window of viability as small as possible, by varying the B cell mutation rate and antibody receptor length. Optimizing these parameters in a physical model leads to values close to the biological ones (82).

### 9.3 The emergence of new strains of a pathogen

Considering a population of viruses at the mean field level leads one to suspect that each individual virus would maximize its reproductive rate, $R_0$. This analysis, however, is approximate because it ignores correlations in the virus population. If there are multiple strains and co-infection of different virus strains in individual hosts, the strains that persist in a population may not be the ones that maximize $R_0$ (83). That is, the virus population will maximize total strain persistence, rather than the replication rate of any individual virus. This result was evident in quasispecies theory, where it is not the bare viral replication rate $f(\xi)$ that is optimized, but rather the renormalized, population-level viral fitness, $f_m$.

## 10. Evolution of the influenza virus

The influenza virus evolves to evade the immune response. This evolution occurs in the population of individuals, and the immune pressure on the virus arises from those individuals who have memory antibodies against the virus due to previous infection or those individuals who have been vaccinated. The antigenic variation of the virus disrupts the antigen-specific immune response (36). This variation arises from point mutation of the RNA that encodes the proteins of the virus that the immune system recognizes as well as recombination and reassortment of the genes of the virus between different strains.

### 10.1 Pressure on the virus to evolve

The virus experiences different pressures that can lead to evolution. First is pressure from the immune system. As discussed in Sections 7.1 and 7.2, the immune pressure from the antibody immune system can be quantified by the $p_{epitope}$ distance between the virus and any strains to which an individual has been previously exposed or immunized. This immune pressure extends to $p_{epitope} < 0.19$ for H3N2 or $p_{epitope} < 0.45$ for H1N1.

There are also pressures on the virus related to its ability to infect and transmit between individuals. For example, typical avian influenzas infect water fowl in the lower respiratory tract, whereas typical human influenzas infect people in the upper respiratory tract. Thus, the influenza virus must alter its binding specificity, from what is known as an α 2,3- to an α 2,6-linkage to sialic acid, when making the transition from birds to humans (84). The ability of the virus to persist in a population also depends on its transmission rate between individuals, and so there is pressure on the virus to achieve a viable transmission rate.

Finally, there are pressures on the influenza virus to survive antiviral treatment. There are a number of antivirals against influenza, including neuraminidase inhibitors and adamantines. Adamantine resistance has evolved rather completely in current influenza strains (85). A single Ser31Asn replacement in the M2 protein is thought to be sufficient to cause resistance. This evolution of the M2 protein has perhaps occurred only once, with that strain then spreading throughout the population, as all adamantine resistant viruses have a shared 17 amino acid replacement signature. The emergence of adamantine resistant viruses also provides further support for the source-sink model with Asia as the source, because there is limited adamantine use in the US and Australia, but a majority of the strains in the US and Australia are resistant (86).

Resistance to the neuraminidase inhibitors is incomplete at present. Neuraminidase inhibitors are 97% efficacious when given 29h after exposure to a susceptible strain (87). These antivirals inhibit the release of newly formed virus particles from infected cells, and the virus is cleared before a systemic infection is established when the inhibitor is given sufficiently early. There are two known mutations that lead to neuraminidase inhibitor resistance: His274Tyr and Asn294Ser (88).

Of note is that while specific amino acid mutations or pairs of mutations are often mentioned as necessary and sufficient for resistance or transmission modification, other mechanisms to achieve these changes are likely possible. For example, more than simply α 2,3- to α 2,6-linkage changes are known to be responsible for human infection (84). In fact, the novel A/H1N1(2009) strain, which seems to be highly infectious and transmissible between humans, is still an α 2,3-linkage binder as of this writing.

### 10.2 The emergence of new influenza strains

Occasionally a strain of influenza that had previously only infected birds or animals will appear in humans. Two recent examples are the H5N1 “bird flu” and the H1N1 “swine flu.” The jumping of a pathogen to a human host is called “zoonosis.” There are three stages (89). The first stage is infection of humans, but no human-to-human transmission. The H5N1 is currently in this stage. The second stage is localized transmission of the new strain among humans. The third stage is sustained transmission among humans which causes an epidemic. The novel A/H1N1(2009) is in this stage.

There are several barriers to achieving the third stage (90). The virus must overcome the human immune system. Additionally, however, the virus may also need to undergo substantial evolution to achieve sustained transmission between humans. It has been said that "in the context of emergence of an influenza A virus strain via a host switch event, it is difficult to predict what specific polygenic changes are needed for mammalian adaptation" (91). In other words, zoonosis is not a predictable event with current theory. In the context of A/H5N1, there has been quite a bit of thought as to what would lead this virus to develop sustained human-to-human transmission (92). For example, A/H5N1 may acquire from reassortment events with A/H1N1 or A/H3N2. Indeed, a significant fraction, up to 25%, of patents infected with one strain of influenza are simultaneously infected with another strain, thus allowing for potential reassortment events (92). It is also of note that consensus sequencing, e.g. as carried out by Roche/454 machines, cannot generally measure evolution within individual hosts. For this reason, published consensus sequences are skewed toward dominant strains within isolates (92).

## 10.3 Point mutation

Point mutation is a fine tuning of the virus once it lands in a new host. To evade the immune system, for example, the hemagglutinin protein must make a sufficiently large modification of $p_{epitope}$ so that immune system memory from prior infection or vaccination does not eliminate the virus. Taking the novel A/H1N1(2009) as an example, the distances between this virus and the A/H1N1 virus from the 2008-2009 season, the 1976 swine flu strain, and the 1918 Spanish flu strain are $p_{epitope}$ = 0.91, 0.28, and 0.38 respectively. Point mutations contributing to other components of the virus fitness, such as transmission, are less well understood at present. For example, "molecular markers predictive of adaptation in humans are not currently present in 2009 A(H1N1) viruses, suggesting that previously unrecognized molecular determinants could be responsible for the transmission among humans" (93).

## 10.4 Reassortment, recombination, and HGT

Segmental reassortment of the 8 coding segments of influenza is a significant contributor to the evolution of novel strains of H1N1. For example, each of the novel H1N1 strains in 1918, 1947, and 1951 came about by reassortment (94), as did the strain in 2009. Multiple lineages of H3N2 co-circulate, persist, and reassort (95).
One study found 14 HA or NA reassortments over 7 years (45). Consistent with this study, it has been estimated that there may be at least 2-3 reassortment events among influenza A strains each year (96). The most common reassortment is incorporation of a novel hemagglutinin or neuraminidase segment to the virus. These are also the proteins undergoing the most rapid point mutation evolution. It appears that the reassortment evolvability of these two proteins has been selected for due to a sufficiently great immune pressure. Reassortment of these segments also occurs at an elevated rate in vitro, in the absence of immune selection, suggesting that the reassortment of these segments may be elevated due to packaging effects.

## 10.5 Novel strain detection

The sequence data of influenza are consistent with large selection pressure (97). While there are long periods of stasis of the dominant strain, this suppression of fluctuations is to be expected in a finite population. Even strong selection pressure, e.g. from the immune system, could lead to a relatively homogeneous virus population during one season but rather different populations in different seasons. The existence of this selection pressure, suppressed as it is by finite population effects, has not been widely recognized (98).

Thus, identification of new strains, before they become dominant, is a significant and non-trivial task. The annual influenza vaccine for the northern hemisphere is chosen in February for the vaccine that will be administered in October. So, novel strains must be identified at least 9 months in advance of when they will be dominant if a vaccine against them is to be identified. As discussed in Section 7, strain classification has been based upon the ferret hemagglutination inhibition assay. However, as discussed in Section 7.2, the correlation between the ferret immune and human immune responses is imperfect. By projecting the strains in the multidimensional sequence space to a two-dimensional sequence space, clusters containing incipient new strains can be identified months to years in advance (99). It appears that novel strains can be identified with clusters containing as few as ten strains.

### 10.6 Detailed simulation of antibody binding

Immune memory from virus in previous years imposes a pressure on the virus to evolve. The pressure is due to antibodies produced in the immune system that chemically bind to the hemagglutinin protein on the virus surface. People who have been vaccinated have antibodies that bind the virus protein $10^2$-$10^3\times$ times as strongly. Since there is pressure on the virus to mutate, when the virus does mutate, this may be to the advantage of the virus. In particular, some mutations may disrupt the antibody/hemagglutinin binding more than others. We postulated that the free energy of the binding is directly correlated to the fitness of the virus (54). That is, we propose that the decrease in the binding constant between antibodies and the mutated hemagglutinin is one of the most significant driving forces of virus evolution.

We have used molecular dynamics (MD) simulation with the CHARMM22 force field to calculate the Gibbs free energy $\Delta\Delta G$ from which the change of binding constant

$$\frac{K_a'}{K_a} = \exp\left(-\Delta\Delta G/RT\right)$$

is obtained, where $K_a'$ and $K_a$ are the binding constant of the mutated complex and the wild type complex, respectively. Thermodynamic integration, with an Einstein crystal reference system for the endpoints, was used to calculate this free energy difference.

We performed calculations on the 1968 H3N2 Aichi strain (100). We noticed that the 10 observed mutations had different levels of stability, and some mutations persisted while others reverted to the original residue one or two years after the first mutation. We found that the mutations which persisted in the human populations were those that most disrupted $\Delta\Delta G$ among the 10 observed. We also calculated the free energy change for a double mutation found in a Guinea pig animal model experiment performed by our collaborators.  This mutation significantly disrupted $\Delta\Delta G$.

### 10.7  Projected dynamics

Coupling the virus evolution within individuals to the transmission of the virus in a population leads to a large system of stochastic dynamics.  Analytic progress is possible when this dynamics is projected to a lower dimension.  For example, stage-structured modeling reduces the number of variables (101).  This approach is a projection of an agent-based approach to be discussed next.

### 10.8  Agent based modeling

Agent based modeling provides a full solution to the stochastic process of virus evolution in a population.  Fluctuation effects are particularly significant to capture effects of strain emergence and extinction (1).  Deterministic modeling, in particular, often leads to an excessive number of strains, with a paucity of extinction (102).

A full agent based model captures the immune pressures on the virus in an individual due to prior exposure or immunization, via $p_{epitope}$ distances (103).  Each individual acquires an immune history specific to its exposure history.  Viruses mutate and evolve in individuals and transmit to other individuals.  So, the variables are the current infection status and immune history of each individual.  Transmission between individuals occurs via social contact, and so the contact network within and between cities is a key parameter in the formulation of the dynamics.

**Figure 10:** The vaccine effect on cumulative attack rate for the initial introduction of two-strains. The multiple-component vaccine (Mc) is superior to the single-component vaccine (Sc) for different fractions of population vaccinated.  After (103).

This detailed model offers specific predictions for the progression of an influenza epidemic and how vaccination may modulate the epidemic (103).  The attack rate in the population, the fraction of the population that becomes infected during a season, is of prime interest and is calculated.   See Figure 10.  The greater the fraction of the population vaccinated, and the earlier the vaccine is administered, the lower the attack rate.  In this way, “herd immunity” is quantified.  In some cases, a multi-strain vaccine is warranted.  For example, in the 2009/2010 season both a seasonal and a novel H1N1 vaccine will be available.  This detailed, agent based approach is the first to quantify the efficacy of a multiple strain vaccine.  The interactions among vaccine components can be non-trivial.  The reports in 2009 of potential negative interactions between the seasonal and the novel H1N1 components of the vaccine, which concerned three provinces in Canada enough to delay administration of the seasonal vaccine until the novel vaccine was ready, are an example (the decision was based on preliminary results – not yet published or reviewed – from four Canadian studies involving 2000 people according to Dr. Don Low at Mount Sinai Hospital) (104).   To further quantify the potential impact of an epidemic, the quantity Population at Risk (PaR) was introduced. The PaR is the maximum expected attack rate, calculated to a 90%, 95%, or 99% confidence interval.

## 11. Evolution of other viruses

There are several patterns into which antigenic variability falls for general pathogens. These patterns are (105)

1) limited variation within host, population wide variation that is independent of space and time (e.g. *S. pneumoniae*)
2) limited variation within host, population wide variation that depends on space and time (e.g. *N. meningitides*)
3) limited variation within host, limited population variation at one time, but rapid population variation over the time scale of years (e.g. Influenza)
4) significant variation in host over time, extensive and increasing diversity in the global population (HIV-1)
5) limited or no antigenic variability, limited diversity in the population, and so nearly complete vaccine or naturally acquired immunity (measles, mumps, rubella)

A strong immune cross reactivity between strains leads to large distances between evolving strains. Conversely, higher transmission, weaker cross immunity, and short-lived infections promote a rapid strain turnover. The prevalence of a pathogen prevale is increased for higher transmissibility for infections with longer duration, or for less immune cross-immunity. Thus, pathogenic diversity is greater for weaker cross immunity and for more transmissible pathogens. A failing of simple epidemic models is that too much diversity is generated too rapidly.

Antigenic diversity is a major challenge to vaccine design, as we saw in Section 7 for the specific case of influenza. For viruses such as HIV or hepatitis C, one wonders whether a vaccine can be protective in a diverse population of hosts (105). For influenza, one wonders whether a universal vaccine, one not requiring annual updates, is possible.

Modeling has shed light on a few features of viral emergence (84). For example, due to their higher mutation rates, new RNA viruses are more likely to emerge than DNA viruses. In addition, zoonosis is more likely from generalist than specialist viruses. Epistatic effects stemming from the necessity to replicate in different hosts, however, disallows much amino acid variation and reduces the speed with which zoonosis occurs. While phylogenetically related species are more likely to have cross species transmission, similar immune systems in related species may suppress virus transmission in new species; detailed modeling could provide insight into this phenomenon.

### 11.1 Evolution of the HIV virus evolution

HAART therapy for HIV has significantly increased the life expectancy of infected patients. The effects on HIV evolution, however, are less clear (106). For example, while HAART increases life expectancy, it also increases viral transmission.

On the other hand, decreased infectiousness decreases the transmission. Thus, mathematical models can determine

1) Optimal use of HAART
2) Epidemiological consequences of HAART and behavioral changes
3) The course of evolution of drug resistance within an individual and within a population
4) Achievable levels of coverage and effectiveness, including effective and efficient use of second line treatments and demographic or health care impacts.

A number of features of HIV evolution have been determined. For example, there is a significant "bottleneck" as the virus transmits from one person to another. The HIV diversity goes down 99% during transmission to new host (107). The evolution of HIV slows down as $CD4^+$ T cells decline, presumably because the immune pressure is reduced (108).

A quasispecies model for HIV evolution demonstrated the emergence of mutants (109). The time scale for a substantial diversity to evolve was 1-2 years. As the mutants emerge, the diversity taxes the capacity of the immune system to respond, due to immunodominance. This mechanism was the first to explain, in a single model, the three phases of HIV. See Figure 11. Moreover, this mechanism suggests a forward-looking vaccine strategy. If a subset of the mutants that are likely to evolve in 1-2 years are vaccinated against during week 14 using the multi-site protocol of Section 8.2, then immunodominance is alleviated, and the immune system is predicted to eradicate the HIV.

**Figure 11:** HIV-1 infection time course, where circles represent the clinical data and the solid line is generated by the model. From (109).

The effects of recombination on HIV have been investigated (110). At moderate population sizes, recombination increased diversity and mean fitness. At larger population sizes, recombination increased fitness but decreased diversity. For small population sizes, we generally expect that recombination may increase fitness due to a reduction of Muller's ratchet effect (111). For larger population sizes, recombination may increase or decrease fitness depending on the sign of epistasis (79, 112).

### 11.2 Evolution of the hepatitis C virus

The evolution of hepatitis C has been quantified after orthotopic liver transplantation (113). Under conditions of mild fibrosis, genetic distance and non-synonymous mutations in the viral quasispecies increased. Under conditions of severe fibrosis, the opposite happened. The interpretation is that under mild disease, there is less pressure on the virus, and a more complex quasispecies can emerge. Conversely, under severe disease, there is substantial pressure on the virus, and a more localized and less complex quasispecies emerge.

### 11.3 Evolution of the dengue virus

Dengue virus has only recently entered the human population, and the virus continues to evolve in the human host. Homologous recombination has been observed in the flaviviridae family to which dengue belongs (114). Recombination within each of the dengue strains has been observed (95). Conclusive proof of recombination between strains has now been obtained from a patient infected with DEN-1 and DEN-2, and with DEN-1,2 recombinants (115). These results illustrate the general importance of recombination to viral evolution.

## 12. Cancer

The immune system controls pre-cancerous cells on a daily basis. Theoretical discussion of cancer and the immune system focuses on cancers of the immune system, immune control of cancer, and potential vaccination strategies.

### 12.1 The mutations leading to cancer

As an example of detailed modeling, a simulation study of human follicular lymphoma B cells has been carried out (116). A model of the germinal center was constructed. A significant number of detailed parameters were identified. The focus of the study was to identify mutations leading to cancerous cells.

### 12.2 Immunodominance

The resistance of cancer to standard therapies has led many researchers to consider immune control. Various effects, including tolerance and immunodominance, however, have limited the ability of the immune system and vaccines to control cancer. These effects occur because cancer is effectively a multi-strain disease, with multiple cancer-associated or cancer-specific epitopes.

Immunodominance is one general mechanism by which cancer cells may escape, either by mutation of the dominant epitope or by loss of the MHC class I allele that expresses the dominant epitope. Cross-presentation of the lost dominant epitope on surrounding cells often sustains the futile immune response. Indeed, many types of cancerous cells are exceptionally adept at evading immune control.

### 12.3 Multi-site vaccination

One means to generate a broader T cell response to a multicomponent cancer vaccine is to administer the different components of the vaccine strains to different physical regions of the patient, in this way breaking the hierarchy of immunodominance (117). These experiments are well explained by the GNK theory of T cell response to multiple antigens (21,118). See Figure 12. Since cancerous cells often express multiple, similar epitopes, this theory could explain a significant fraction of the immunobiological responses to cancer.

**Figure 12:** Specific lysis for a two component vaccine with two completely different cancer epitopes: a) single site injection b) dual site injection with each epitope in the vaccine draining to a different lymph node. From (21).

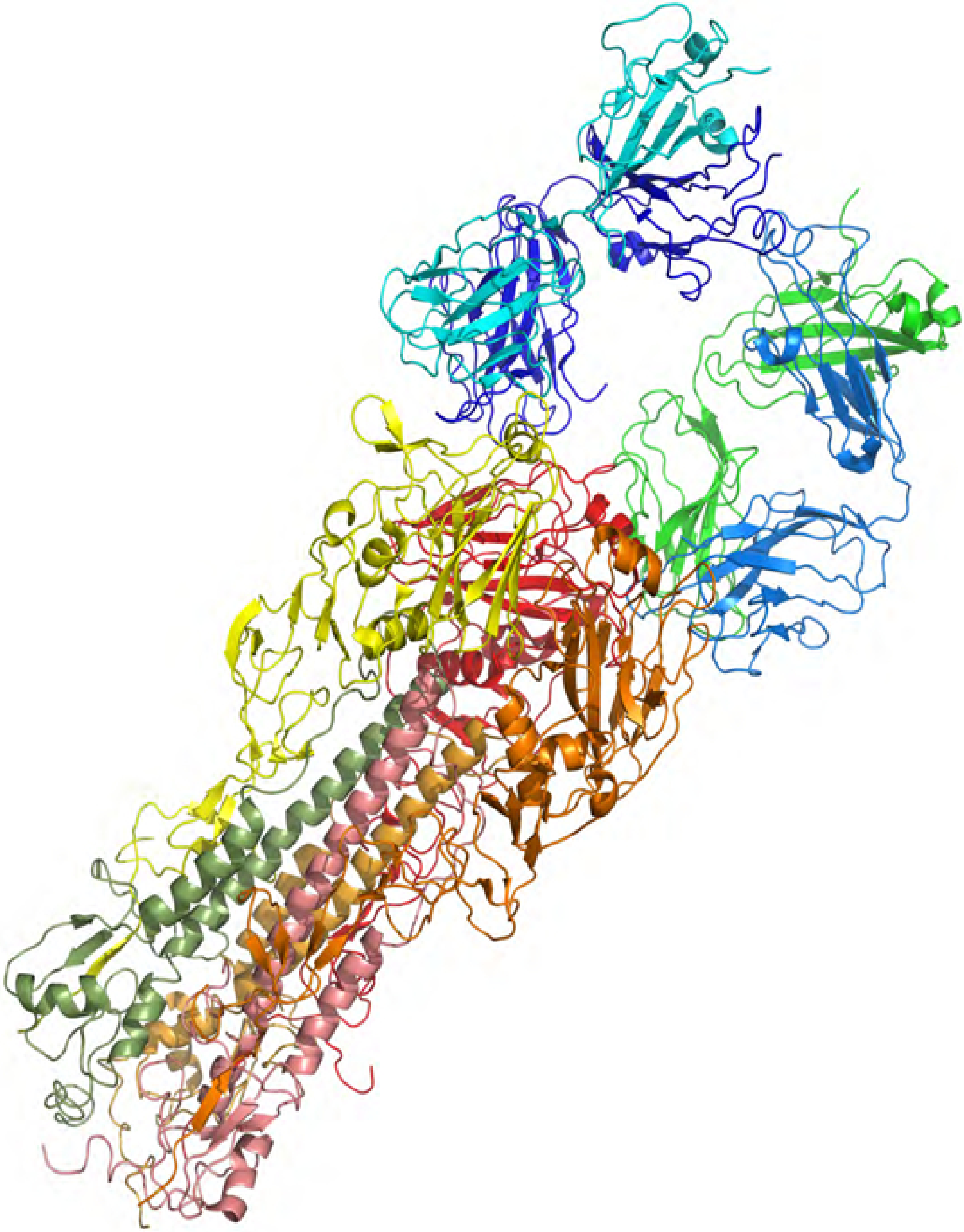

Figure 1

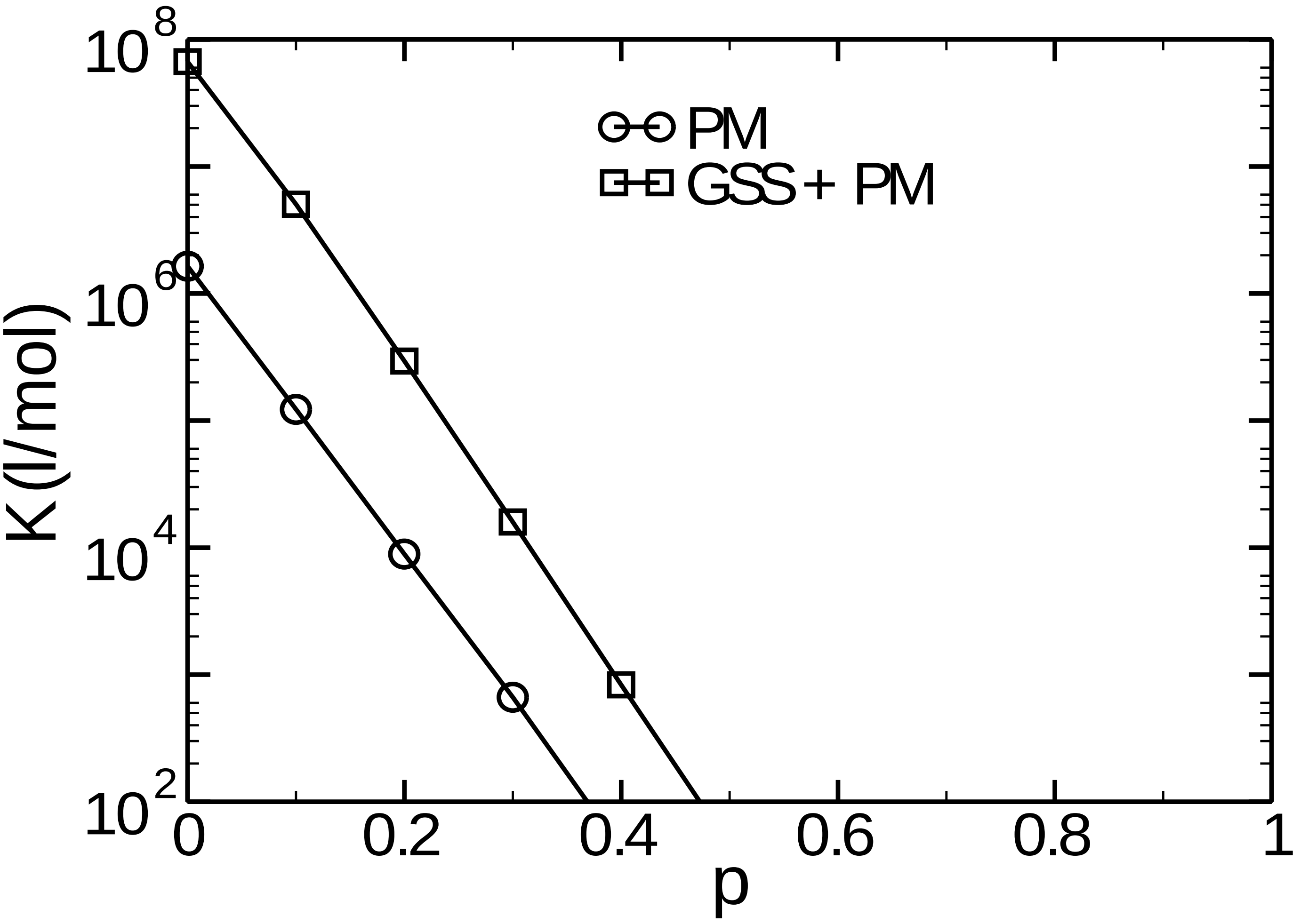

Figure 2

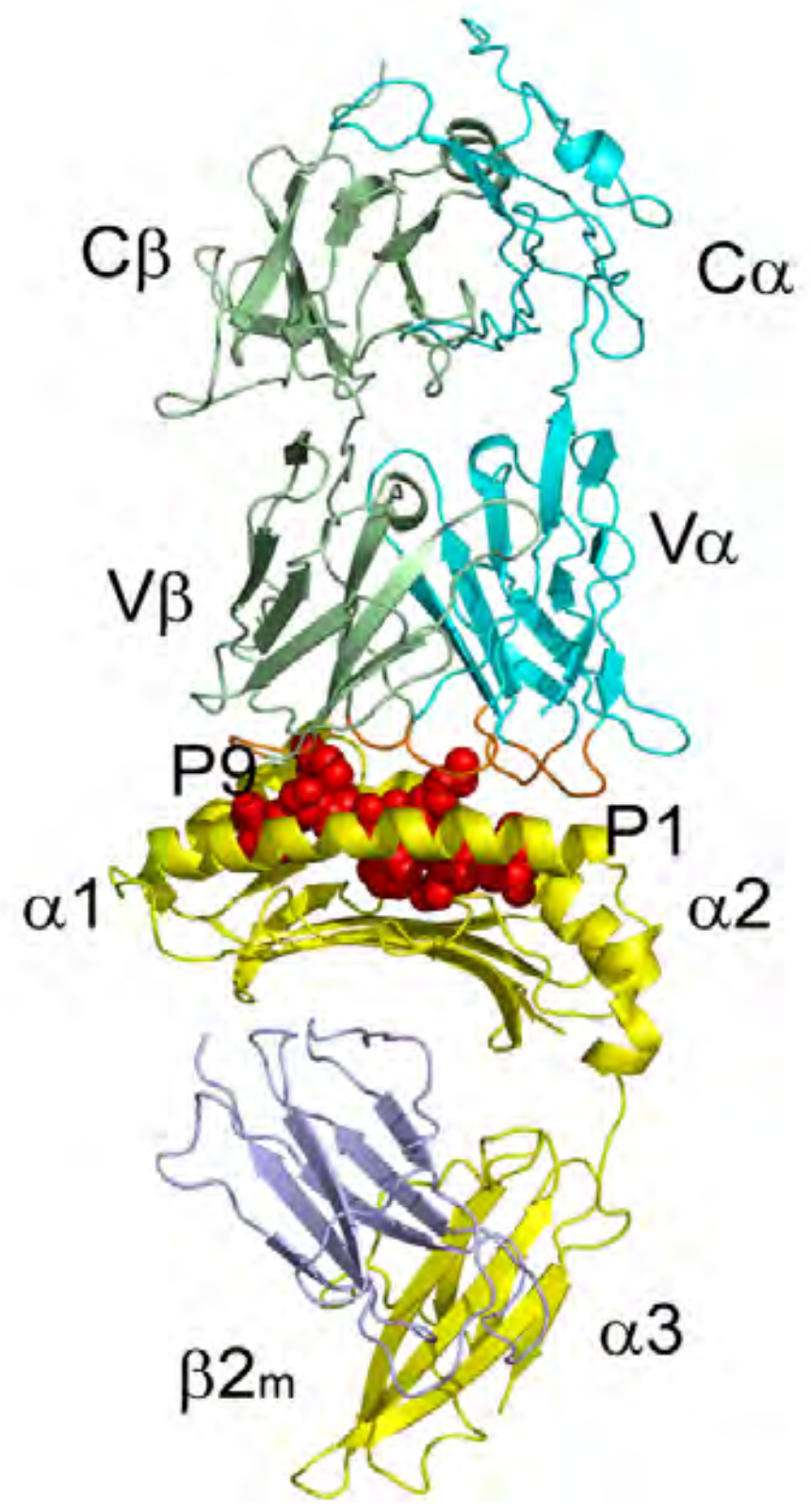

Figure 3a

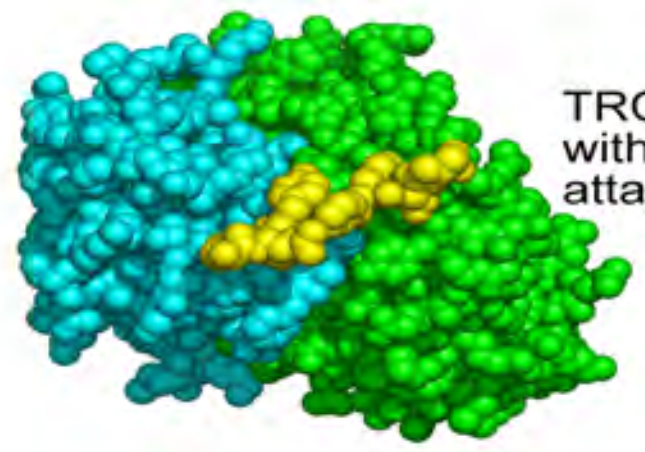

Figure 3b

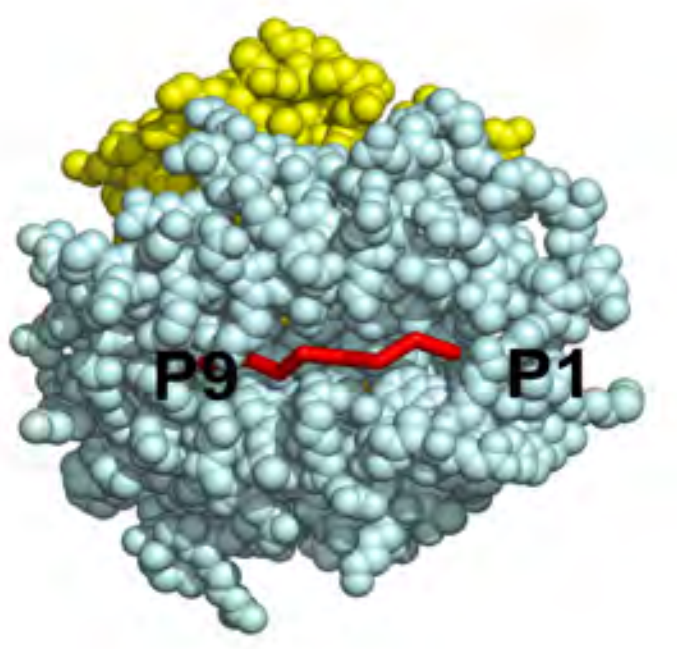

Figure 3c

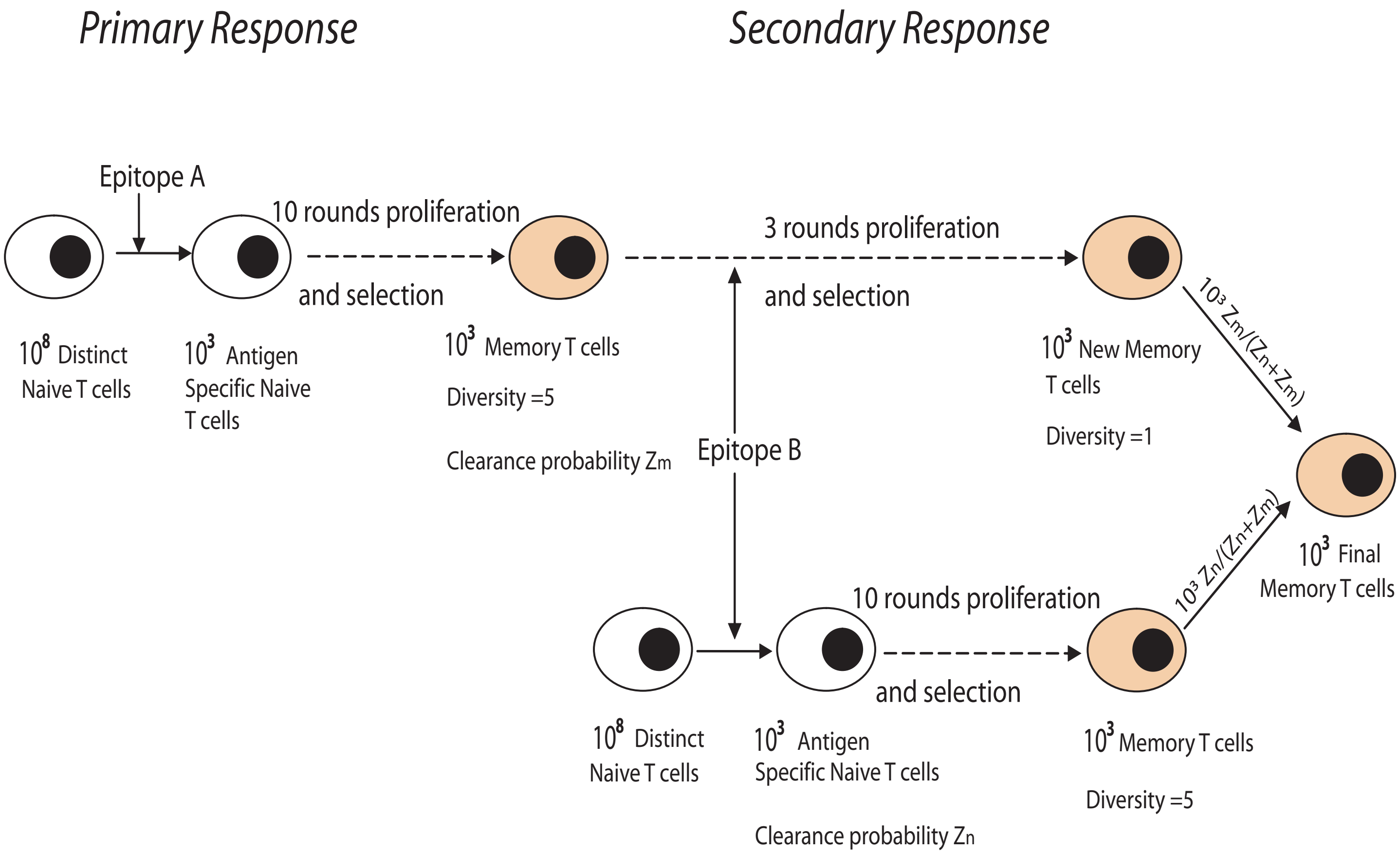

Primary Response
Secondary Response
Epitope A
10 rounds proliferation
and selection
3 rounds proliferation
and selection
10^8 Distinct Naive T cells
10^3 Antigen Specific Naive T cells
10^3 Memory T cells
Diversity =5
Clearance probability Zm
Epitope B
10^3 New Memory T cells
Diversity =1
10^3 Zm/(Zn+Zm)
10^3 Zn/(Zn+Zm)
10^3 Final Memory T cells
10 rounds proliferation
and selection
10^8 Distinct Naive T cells
10^3 Antigen Specific Naive T cells
10^3 Memory T cells
Diversity =5
Clearance probability Zn

Figure 4

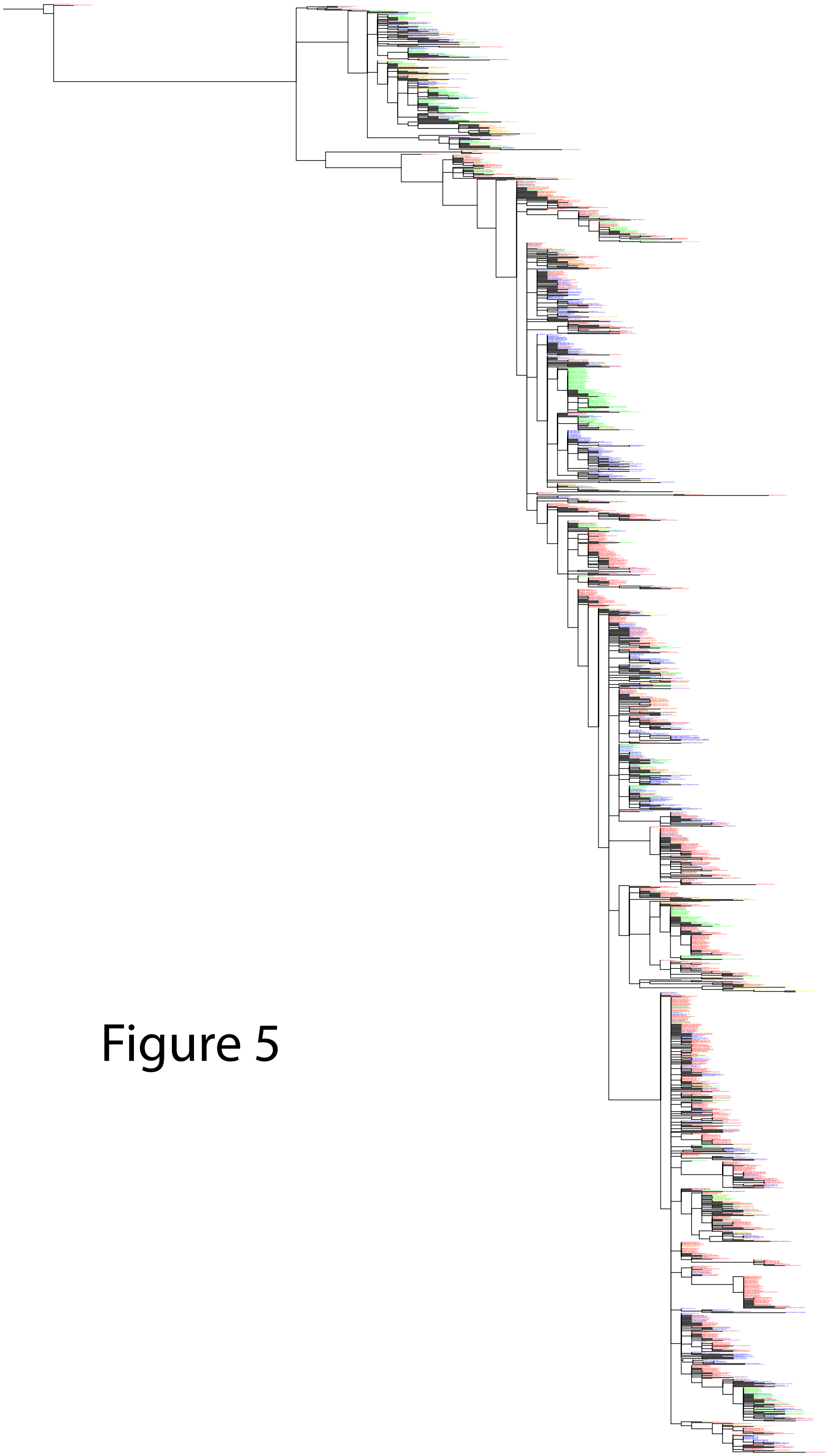

Figure 5

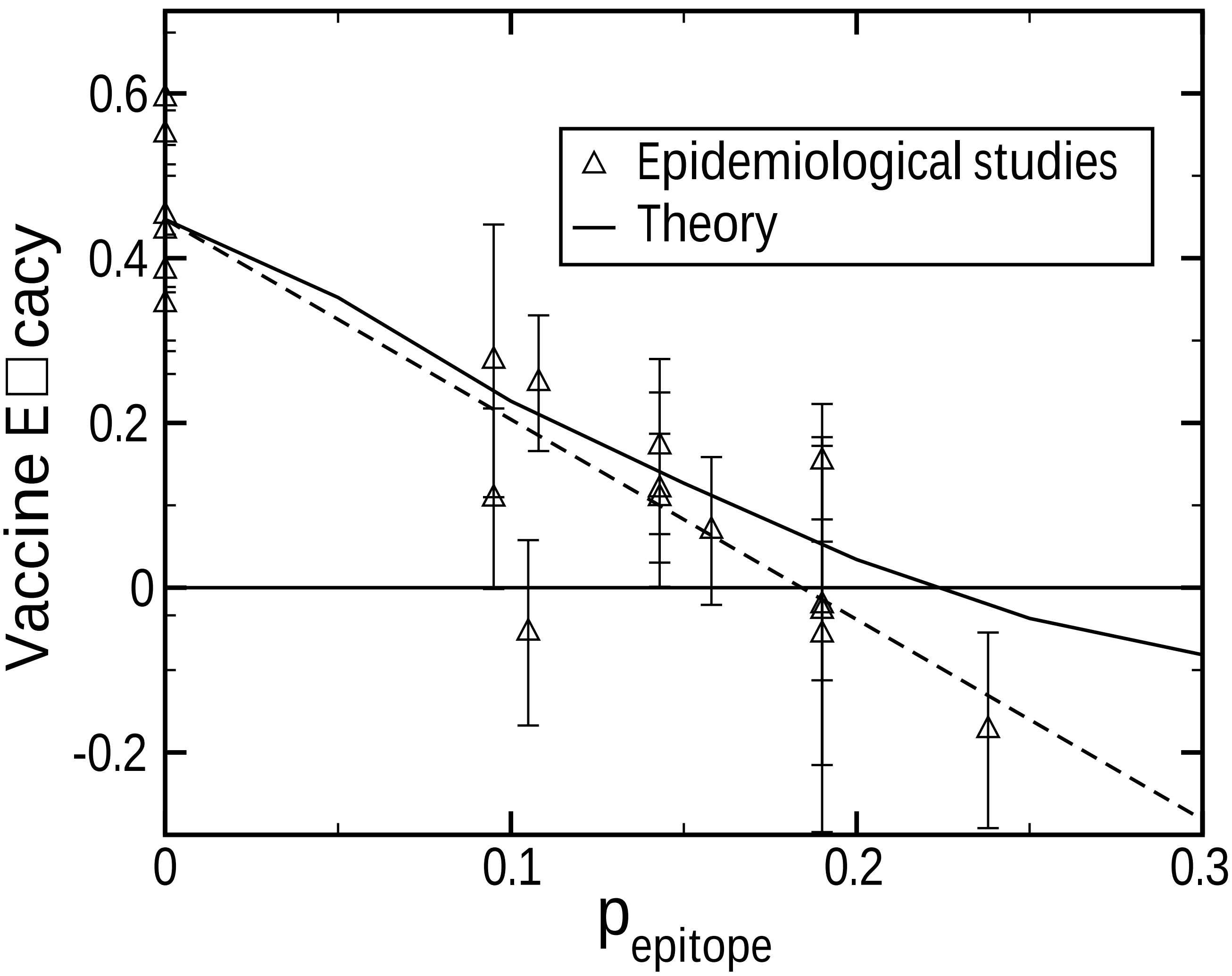

Figure 6

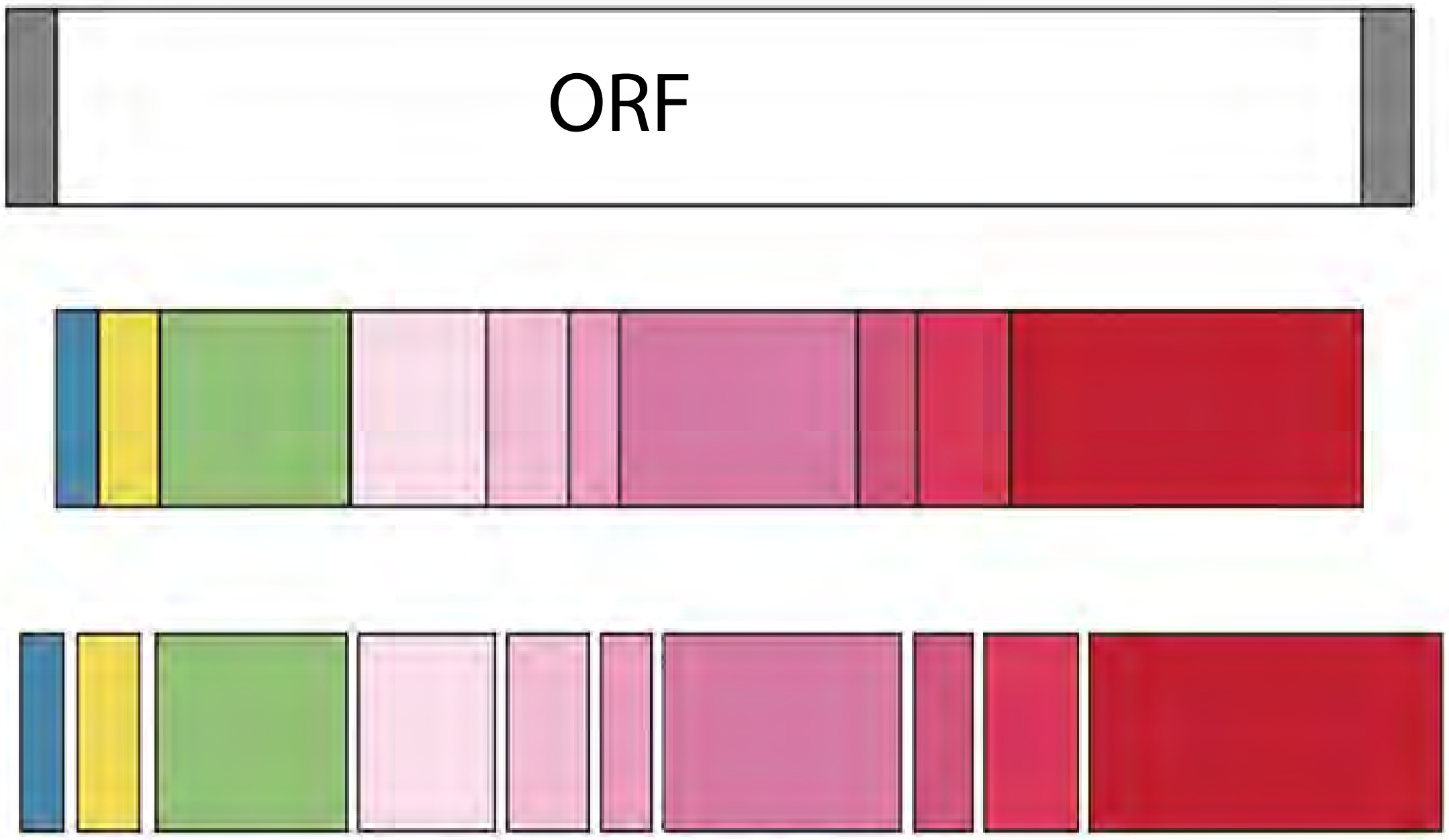

| Response | C | pM | E | NS1 | 2a | 2b | NS3 | 4a | 4b | NS5 |
|---|---|---|---|---|---|---|---|---|---|---|
| Antibody | | + | +++ | ++ | | | | | | |
| T cell | + | + | + | + | | | +++ | | + | + |

Figure 7

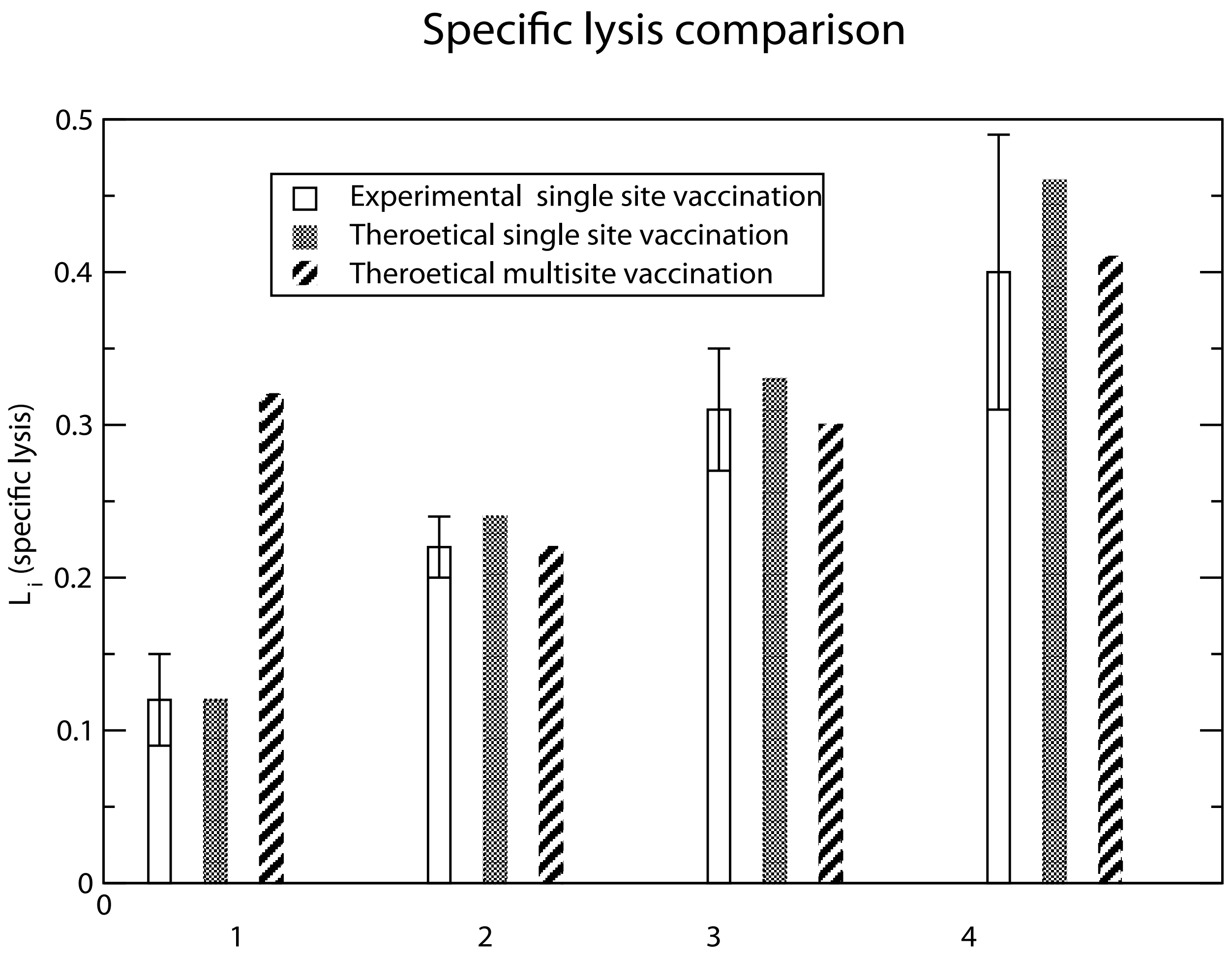

Figure 8

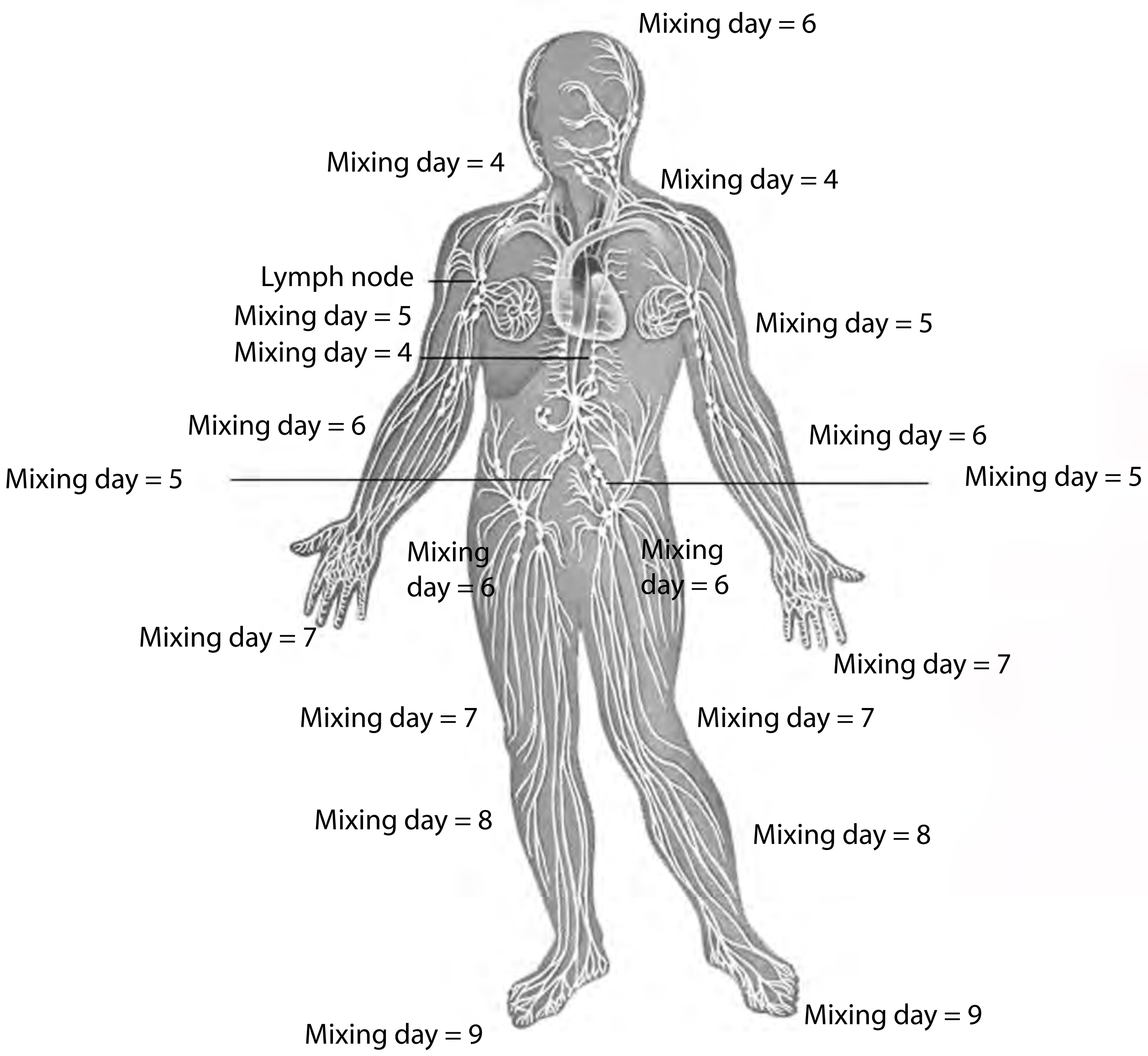
Mixing day = 6
Mixing day = 4
Mixing day = 4
Lymph node
Mixing day = 5
Mixing day = 5
Mixing day = 4
Mixing day = 6
Mixing day = 6
Mixing day = 5
Mixing day = 5
Mixing day = 6
Mixing day = 6
Mixing day = 7
Mixing day = 7
Mixing day = 7
Mixing day = 7
Mixing day = 8
Mixing day = 8
Mixing day = 9
Mixing day = 9

Figure 9

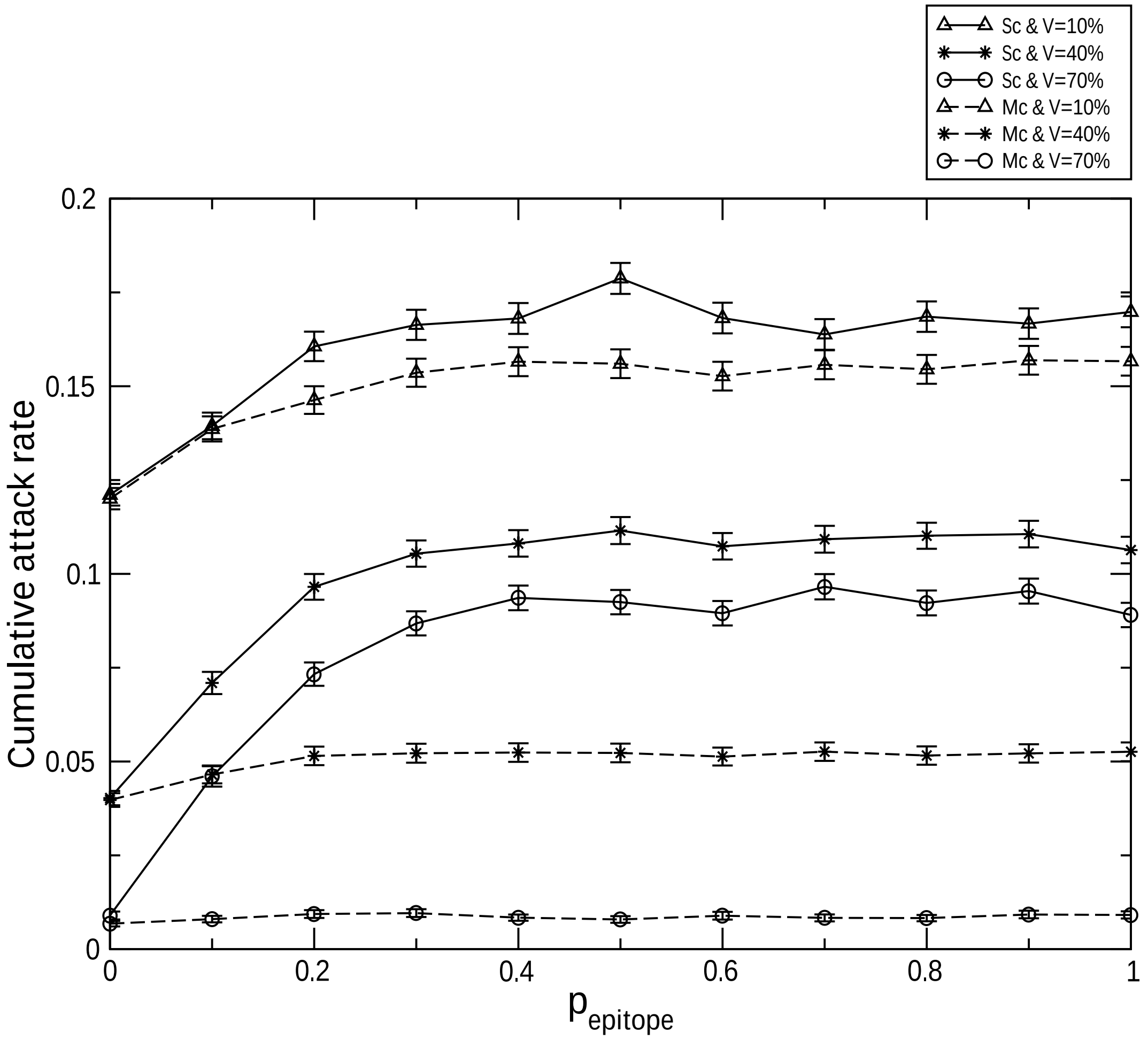

Figure 10

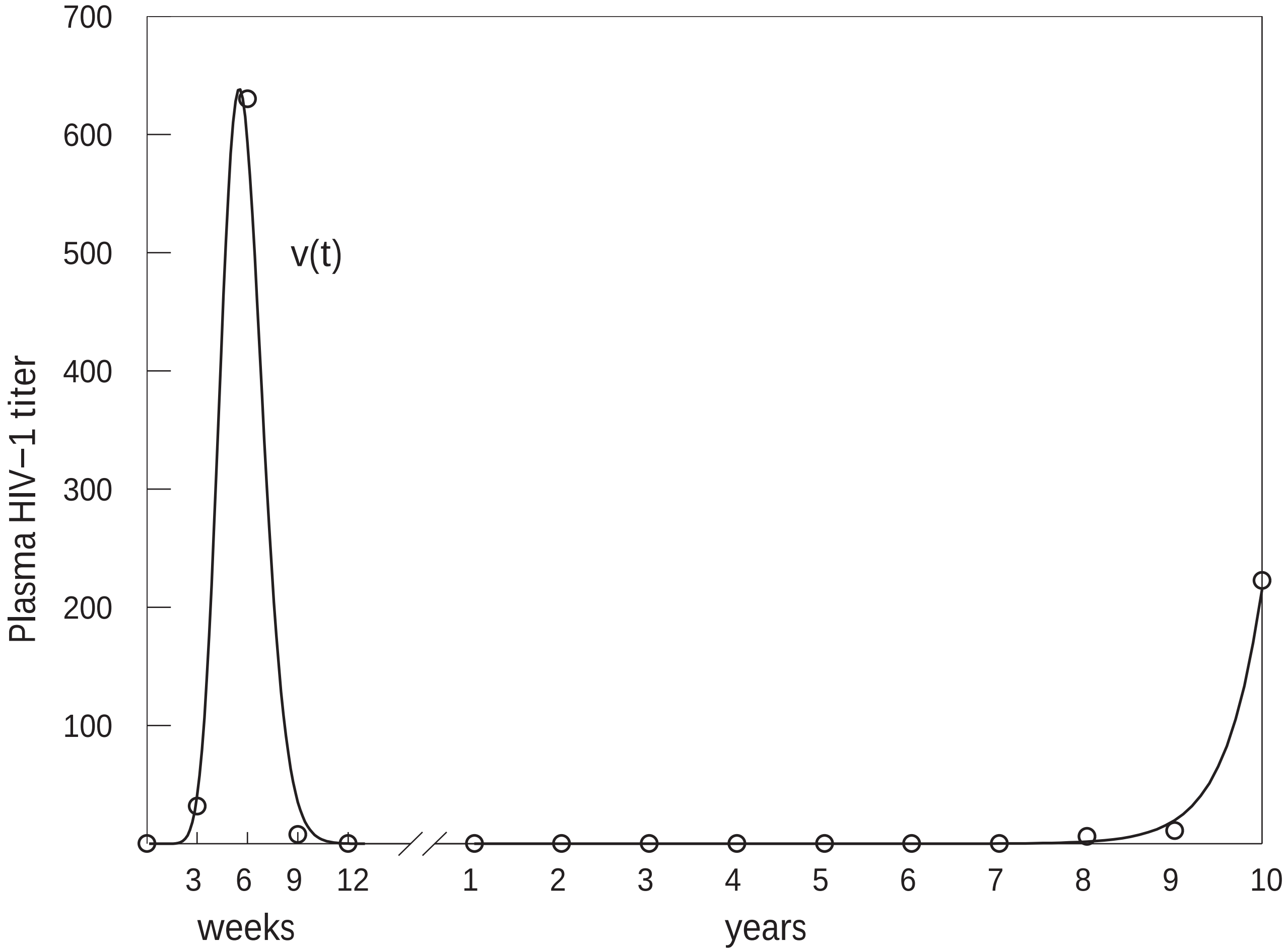

Figure 11

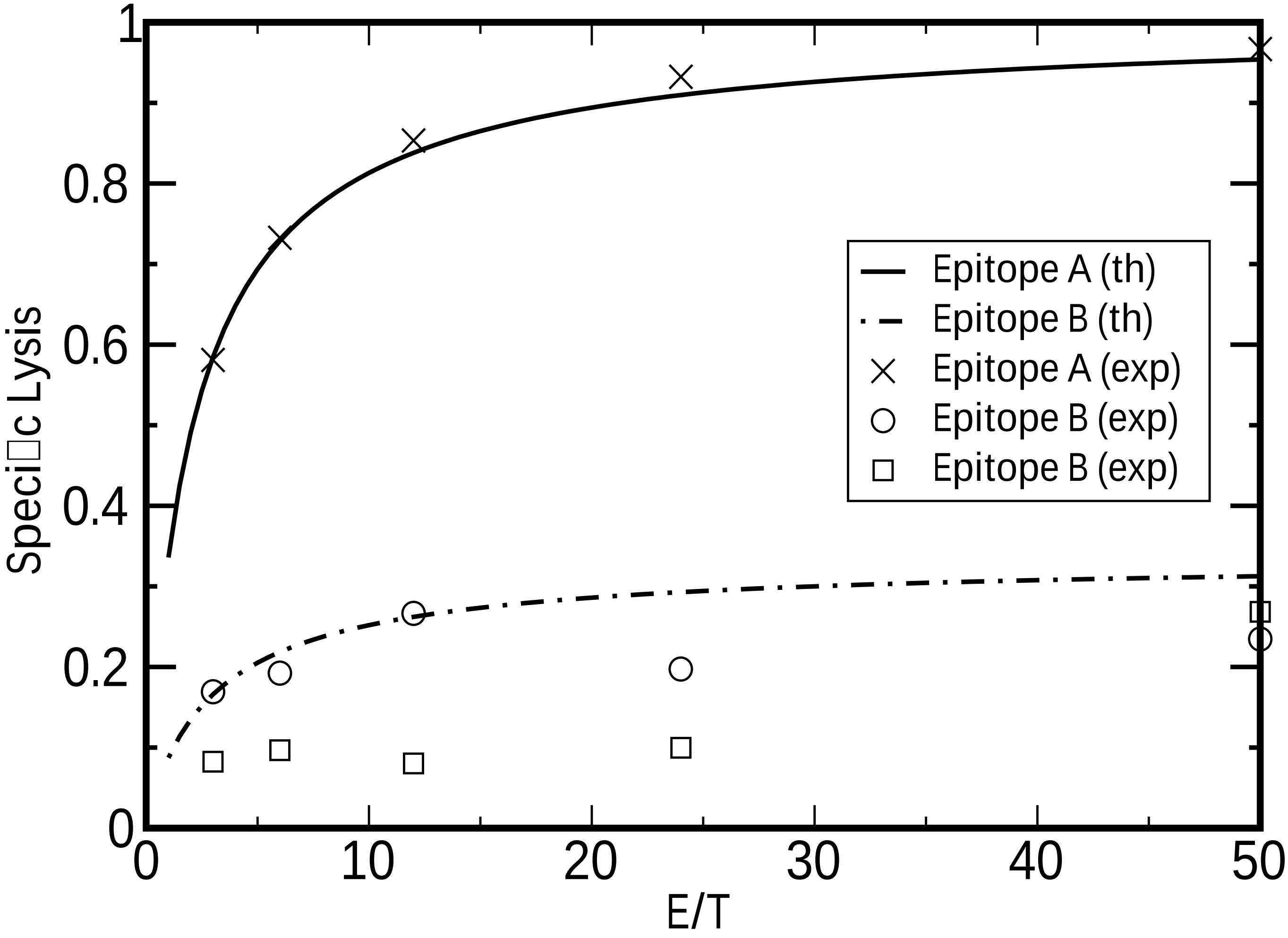

Figure 12a

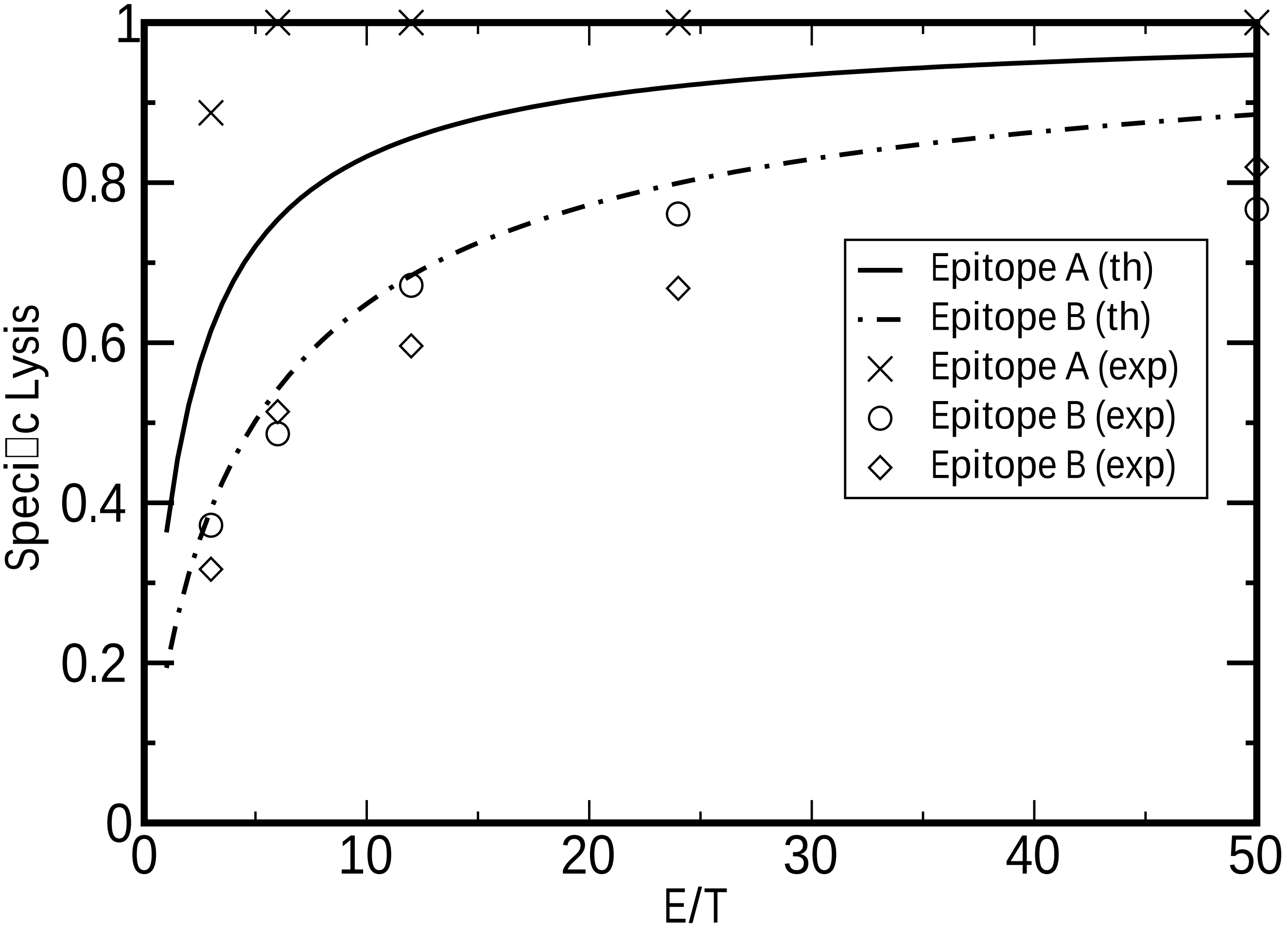

Figure 12b